\documentclass[final,authoryear,5p]{elsarticle}

\biboptions{authoryear}

\usepackage{graphicx}
\usepackage{float}
\usepackage{amsmath}
\usepackage{flushend}

\usepackage[ps2pdf,%
a4paper=true,%
breaklinks=true,%
colorlinks=true,%
pdfauthor={Sreehari et al.},%
pdftitle={Constraining the mass of the black hole GX 339-4 }%
]{hyperref}

\journal{Advances in Space Research}

\begin{document}

%%%%%%%%%%%%%%%%%%%%%%%%%%%%%%%%%%%%%%%%%%%%%%%%%%%%%%%%%%%%%%%%%%%%%%%%%%%%%
%% Frontmatter
\begin{frontmatter}


% Use the tnoteref command within \title and fnref within \author or \address for footnotes;
% use the corref command within \author for corresponding author footnotes;
% \author{Name\corref{cor1}\fnref{label2}}
% \address{Address\fnref{label3}}

\title{Constraining the mass of the black hole GX 339-4 using spectro-temporal analysis of multiple outbursts \tnoteref{footnote1}}
%\tnotetext[footnote1]{This template can be used for all publications in Advances in Space Research.}

% Use optional labels to link authors explicitly to addresses:

%\author{Sreehari H.\corref{cor}\fnref{footnote2}}
\author{Sreehari H.\corref{cor}}
\address{Space Astronomy Group, ISITE Campus, URSC, Outer Ring Road, Marathahalli, Bangalore, 560037, India\\
         Indian Institute of Science, Bangalore, 560012, India}
\cortext[cor]{Corresponding author}
%\fntext[footnote2]{}
\ead{hjsreehari@gmail.com}

% Url can be given like this:
% \ead[url]{http://www.elsevier.com/wps/find/authorsview.authors/latex}

%\author{N. Iyer\corref{cor1}}
\author{Nirmal Iyer}
%\author{N. Iyer\fnref{footnote3}}
\address{Albanova University Centre, KTH PAP, Stockholm, 10691, Sweden}
%\fntext[footnote3]{Additional information about the second and third authors}
%\cortext[cor1]{}
\ead{nirmal.iyer@gmail.com}

\author{Radhika D.}
\address{Department of Physics, Dayananda Sagar University, Bangalore, 560068, India}
%\fntext[footnote4]{Additional information about the co-authors}
\ead{radhikad.isac@gmail.com}

\author{Anuj Nandi}
\address{Space Astronomy Group, ISITE Campus, URSC, Outer Ring Road, Marathahalli, Bangalore, 560037, India}
\ead{anuj@isac.gov.in}

\author{Samir Mandal}
\address{Indian Institute of Space Science and Technology, Trivandrum, 695547, India}
\ead{sam.cenb@gmail.com}

\begin{abstract}
We carried out spectro-temporal analysis of the archived data from multiple outbursts spanning over the last two decades from the black hole X-ray binary GX 339-4. In this paper, the mass of the compact object in the X-ray binary system GX 339-4 is constrained based on three indirect methods. The first method uses broadband spectral modelling with a two component flow structure of the accretion around the black hole. The broadband data are obtained from {\it RXTE (Rossi X-ray Timing Explorer)} in the range 3.0 to 150.0 keV and from {\it Swift} and {\it NuSTAR (Nuclear Spectroscopic Telescope Array)} simultaneously in the range 0.5 to 79.0 keV. In the second method, we model the time evolution of Quasi-periodic Oscillation (QPO) frequencies, considering it to be the result of an oscillating shock that radially propagates towards or away from the compact object. The third method is based on scaling a mass dependent parameter from an empirical model of the photon index ($\Gamma$) - QPO ($\nu$) correlation. We compare the results at 90 percent confidence from the three methods and summarize the mass estimate of the central object to be in the range $8.28 - 11.89~ M_{\odot}$. 
%{\bf From an analysis based on probability density functions, we also deduce the most likely mass of 10.10 - 10.32 $M_{\odot}$.}

\end{abstract}

\begin{keyword}
%first keyword \sep second keyword \sep more keywords
%first keyword; second keyword; more keywords
Black hole candidate; accretion discs; X-ray outbursts
\end{keyword}

\end{frontmatter}

\parindent=0.5 cm

%%%%%%%%%%%%%%%%%%%%%%%%%%%%%%%%%%%%%%%%%%%%%%%%%%%%%%%%%%%%%%%%%%%%%%%%%%%%%
%% Main text
\section{Introduction}

Black Hole (BH) X-ray binaries (XRBs) are very interesting objects to study because of their peculiar behaviour like sudden increase in intensity leading to an outburst. Such outbursts typically result in the source undergoing spectral state transitions occupying the low-hard state (LHS),  
hard-intermediate state (HIMS), soft-intermediate state (SIMS) and the high-soft state (HSS) during the rising phase and a reverse trend during the decay phase \citep[and references therein]{2001Homan, belloni2005evolution, nandi2012accretion}. These states are characterized by the variations in spectral and temporal properties as the outburst progresses. 
The Hardness-Intensity Diagram (HID) \citep{Macncop2003, HoBel2005, RemiMc2006, nandi2012accretion, radhika2014spectro, Nandi2018} is the log-log plot of hardness (ratio of flux in higher energy band to the flux in lower energy band) versus intensity in the total energy band. The HID for an entire outburst usually takes the shape of a `q' and hence is often referred to as the `q'- diagram. This diagram helps in identifying spectral state transitions of a BH XRB. These state transitions give rise to evolution of spectral parameters obtained from various models,  which can be studied to get a better understanding of accretion dynamics around a black hole. 

It is generally accepted that there is a disc along the equatorial plane of the black hole \citep{ShaSu1973} and around the black hole there is a `corona' or `Compton cloud' \citep{SuTi1980} which inverse-Comptonize soft radiation from the disc. Combining these, several disc-corona models \citep{Svensson1994, Zycki1995, Done2006, Ingram2011} have been proposed to explain radiations from black hole binary systems. These models have helped to understand in detail the evolution of energy spectral parameters like disc temperature and photon index during the outburst. Model with two component accretion flows \citep{CT95,Smith2002,Wu2002} is also considered often. In the two component accretion flow model \citep{CT95}, the BH is considered to be surrounded by a post-shock corona (PSC) and apart from the Keplerian disc there is a sub-Keplerian halo as well. This model is used for broadband spectral modelling to determine accretion rates and black hole mass \citep{iyer2015determination, Nandi2018, Rad2018}. 

The temporal properties of BH sources have given clear evidence for the occurrence of low frequency QPOs in the range 0.1 - 20 Hz. These are further divided into 3 types i.e. A, B and C based on the value of QPO frequency, width and amplitude \citep{Casella2004, belloni2005evolution}. It has been found that for most of the outbursting sources the value of QPO frequency increases with time during the rising phase of the outburst, gets saturated during the SIMS and decreases as the decay phase occurs \citep{belloni2005evolution, nandi2012accretion}. C-type QPOs in particular go through a monotonic increase in frequency during the rising phase of the outburst. They show the reverse trend during the decay phase of the outburst. The C-type QPO frequency is related to the size of the oscillating region \citep{Chakra2008}. From this it is inferred that, during rising phase as the QPO frequency increases the size of the oscillating region decreases and vice versa during the decay phase. The temporal evolution of C-type QPOs have been modelled by \cite{Chakra2008} with black hole mass as a parameter. Interestingly, a positive correlation has also been observed for the variation of QPO frequency with the spectral photon index ($\Gamma$) which keeps increasing in the rising phase of the outburst and gets saturated as the source enters its SIMS \citep{TF04}. This correlation has been empirically modelled by \cite{shaposhnikov2007determination} (hereafter ST07) and used to obtain source mass based on a scaling relation. 

In this paper, we have considered the well known BH source GX 339$-$4 during its 2002, 2004, 2007 and 2010 outbursts in the era of {\it RXTE} \footnote{https://heasarc.gsfc.nasa.gov/docs/xte/XTE.html} and the 2013 and 2015 outburst observed by {\it Swift} \footnote{http://www.swift.ac.uk/} and {\it NuSTAR} \footnote{https://www.nustar.caltech.edu/}. The source is also known as 1H J1659$-$487 \citep{blackcat}.  
Most of the outbursts of GX 339-4 lasted for a duration of a year or more \citep{SreeJAA2018}. The evolution of spectral and temporal properties of this source during the 2002 outburst have been remarkable and paved way for having a clear understanding of the HID \citep{belloni2005evolution}. Identification and classification of the different types of QPOs during the RXTE era have already been done by \cite{belloni2005evolution, motta2011low}. 

\cite{blackcat} has provided the photometric properties determined for this source till date. Though many detailed observations have been done on the system (\cite{samimi1979gx339}, \cite{grebenev1991detection}),  a constraint on the mass of the central BH is yet to be made. The fact that GX 339$-$4 lies very much within the  galactic disc makes it difficult to dynamically estimate its mass, as even in quiescent state the signature of companion is hardly detectable. Nevertheless attempts to estimate the mass of the source  by \cite{hynes2003dynamical} based on Na-I and Ca-II line emissions resulted in a lower mass limit of $5.8 \pm 0.3 ~M_{\odot}$. Further \cite{munoz2008masses} has come up with a tighter bound of minimum $6~ M_{\odot}$ based on a \lq stripped giant\rq $~$ model where in the companion mass is assumed to be $ \geq 0.17~M_{\odot} $. Later, \cite{shaposhnikov2009determination} (hereafter ST09) has estimated the mass of the black hole in GX 339-4 to be $12.3 \pm 1.4~M_{\odot}$ based on a photon index-QPO scaling method. Also, \cite{parkermass2016} has estimated the mass of the black hole to be $9.0^{+1.6}_{-1.2}~M_{\odot}$ and a distance of $8.4 \pm 0.9 ~kpc$ by spectral modelling of {\it XRT} and {\it NuSTAR} data. But \cite{parkermass2016} have used only the data from the very high state of GX 339-4. Recently, \cite{Heida17} determined the mass function to be $f(M)=1.91 \pm 0.08~M_{\odot}$ and a mass range of $2.3-9.5~M_{\odot}$ at a binary inclination of $37^{\circ}$ to $78^{\circ}$. From the above mentioned studies, the mass of the compact object of GX 339-4 is gauged to be in a very wide range from about 2.3 $M_{\odot}$ to 13.7 $M_{\odot}$. In this paper, we attempt to give a better constraint on the mass of the black hole GX 339-4.

Using spectral and temporal modelling, mass of the sources IGR J17091$-$3624 and XTE J1859$+$226 have been estimated previously by \cite{iyer2015determination} and  \cite{Nandi2018} respectively. In this paper, we explore the spectral and temporal evolution of GX 339$-$4 during all its outbursts in the RXTE era. We study the variation of QPO frequency with time and the correlation between spectral photon index and QPO frequency in order to estimate the source mass. We also model the X-ray energy spectra from RXTE using the two component flow model. Further as RXTE observations are not available after 2012, we analyse the broadband spectra from simultaneous {\it Swift-XRT} and {\it NuSTAR} observations for the outbursts in 2013 and 2015. Finally, combining all these results we provide a mass range for the compact object in GX 339-4. 

The paper is organized as follows: In the next section we discuss the observations and data analysis procedures using the software HEAsoft \footnote{https://heasarc.nasa.gov/lheasoft/}. In \S ~\ref{methods}, we explain the different methodologies used to indirectly estimate the mass of the compact object in a black hole binary using spectral and temporal data. In \S ~\ref{results}, we present the results of applying these methods on the observational data from the source GX 339-4. In \S ~\ref{disc} we discuss on results and mention the caveats of the methods used. Finally, we provide concluding remarks on the results in \S ~\ref{conclude}.

\section{Observation and Analysis}

We have used High Energy Astrophysics Science Archive Research
Center (HEASARC) archival data to analyse the outbursts of GX 339$-$4 in the period 1997-2015. We analysed the 2002, 2004, 2007 and 2010 outbursts of GX 339-4 using \textit{RXTE} data and 2013 and 2015 outburst using simultaneous observations from \textit{Swift} and \textit{NuSTAR}. The 1998-1999 outburst was partially observed by \textit{RXTE} and so the data is insufficient for modelling. The spectral and temporal analysis are performed using HEAsoft v 6.21. 
\subsection{\textit{RXTE}}
The RXTE consists of an All Sky Monitor (ASM), Proportional Counter Array (PCA)  and  High-Energy X-ray Timing Experiment (HEXTE) instruments. We use data from PCA and HEXTE for the analysis presented in this paper. PCA and HEXTE together covers a broad energy range extending from 3.0 keV to 150.0 keV. To obtain the PCA spectral data, at first we generate the good time interval (gti) file using {\it maketime}, and then apply this gti and filter file information into {\it saextrct}. The $cmbrightVLE$ background model and the South Atlantic Anomaly (SAA) history file for PCA are given as inputs to the {\it pcabackest} command to estimate background. Further the {\it saextrct} command is used to extract the background spectrum for PCA. The $pcarsp$ tool is used to generate the response matrix. Of the five proportional counter units (PCUs) of PCA, we used only PCU2 which was stable and in working condition over all considered outbursts.

The HEXTE instrument has two clusters A and B. We use cluster A data for our analysis.
The source spectrum was generated using {\it saextrct} and the background spectrum using {\it hxtback} followed by {\it saextrct}. For the 2007 outburst, we used the HEXTE Cluster B data as the cluster A experienced a rocking anomaly where the on and off source modulation ceased \footnote{https://heasarc.gsfc.nasa.gov/docs/xte/whatsnew/big.html}. For the 2010 outburst as the satellite had a rocking problem with cluster B also, the mission team decided to have cluster B permanently in off-source position and cluster A in on source position. In this case, the source spectrum was extracted from FS50* files and the corresponding background was generated from FS56* files which were further processed with $hxtebackest$ to obtain background spectrum corresponding to the FS50* files. The deadtime correction was done using $hxtdead$ and the HEXTE response was generated with $hxtrsp$ (see \cite{Rad2016rxte} for details). The spectral modelling was done using XSPEC-12.9.1. For broadband spectral modelling, we used \textit{RXTE}'s PCA and HEXTE combined spectrum. We used 1$ \% $ systematic error for the spectral analysis with {\it RXTE} data.

For the temporal data, we consider the science event files and made use of the tool {\it seextrct} applying a bin time of 8 ms to generate the lightcurves. The Power Density Spectrum (PDS) for each data set corresponding to 3.0 - 30.0 keV was obtained from the lightcurves using the {\it powspec} tool. These power spectra were modelled with {\it powerlaws, constants and Lorentzians}. We extracted QPO parameters for the Lorentzians  that had a quality factor $\nu/FWHM$ greater than 3, where $\nu$ is the Lorentzian centroid and $FWHM$ is the Lorentzian width. The QPO parameters  obtained are consistent with \cite{belloni2005evolution} and \cite{motta2011low}. 

\subsection{\textit{Swift} and \textit{NuSTAR}}
As RXTE was decommissioned in 2012, we searched for broadband data from other instruments for the period after 2012. So for the 2013 and 2015 outbursts of GX 339-4, we used data from {\it Swift-XRT} \citep{Evans2009}  and data from {\it NuSTAR} observatory to do the simultaneous broadband spectral analysis. The {\it Swift} spectral data were obtained using standard procedures \footnote{http://www.swift.ac.uk/user\_objects/index.php} in the energy range 0.5 to 10.0 keV, which is then grouped to a minimum of 30 counts per bin using the \textit{grppha} tool. 

With the \textit{NuSTAR} data, we initially ran the \textit{nupipeline} tool to generate the pipeline outputs. The source and background regions each of 30$^{\prime \prime}$ were chosen from the cleaned event file using the {\it ds9} tool. Further, we executed the \textit{nuproducts} command to obtain the energy spectra in the range 3.0 to 79.0 keV, grouped at a minimum of 30 counts per bin. Combining \textit{XRT} and \textit{NuSTAR} for simultaneous observations, we obtain broadband spectra ranging from 0.5 keV to 79.0 keV.

\section{Methodology}
\label{methods}

In this paper, we apply three different methods to estimate the mass of the black holes in X-ray binary systems. The first method is based on the two component accretion flow model. The second and third methods involve observations of QPOs, one based on temporal evolution of QPOs and the other depends on QPO frequency correlation with the photon index obtained from spectral modelling.

\subsection{Method I: Two Component Accretion Flow} 

We consider an accretion disc around a Schwarzschild black hole. The accretion disc consists of two components \citep{CT95}, one being the Keplerian disc at the equatorial plane and the other a sub-Keplerian halo above and below the Keplerian disc. The in-falling matter necessarily becomes sub-Keplerian \citep{GiriSKC2013} near the black hole horizon to satisfy the inner boundary conditions.
A sub-Keplerian flow with multiple sonic points, becomes supersonic after crossing the outer sonic point and connect the inner sonic point through a shock transition \citep{Das2001} where the centrifugal force becomes comparable to gravity. The Keplerian disc remains subsonic \citep{Das2001} throughout and it is truncated at the shock. The region between the shock and the black hole horizon is the post shock corona (hereafter PSC) which is sub-Keplerian in nature. At the shock the supersonic flow suddenly becomes subsonic and the excess kinetic energy is converted into thermal energy. As a result the post-shock region becomes hotter and denser than the pre-shock region. The soft X-rays emitted by the X-ray binary system corresponds to thermal emission from the Keplerian disc. This can be modelled with a multi-temperature blackbody like {\it diskbb} model available in XSPEC. Inverse-Comptonization of the soft photons from the optically thick Keplerian disc by the hot electrons in the post shock corona gives rise to a {\it powerlaw} distribution of hard X-rays \citep{TitLyu1995,CT95}. In the two component accretion flow model, the contribution from the soft X-rays are accounted for by the Keplerian-disc accretion rate and the contribution from hard X-rays varies with the accretion rate of the sub-Keplerian component. In this method, mass of the source is determined by spectral modelling. In calculating the radiation spectrum, the mass of the source decides the soft photon flux from the Keplerian disc and the number density of electrons in the central cloud (Compton corona i.e., PSC) through shock location.   

The two component accretion flow is included in XSPEC as an additive table model \citep{iyer2015determination}. This model has five parameters namely the black hole mass, halo accretion rate, disc accretion rate, shock location and normalization. The accretion rates are in units of Eddington rate, shock location is in units of $r_g=2GM/c^2$ (Schwarzschild radius) where $G$ is the gravitational constant and $c$ is the speed of light and black hole mass is in units of $M_{\odot}$. The normalization varies proportional to $cos(\theta)/D^2$ where $\theta$ is the angle of inclination of the system in degrees and $D$ is the distance to the source in units of 10 kpc. Considering a distance of 8.4 kpc and an inclination angle of $37^{\circ}$ to $78^{\circ}$ for GX 339-4, we get a norm value ranging from 0.58 to 2.26. For our analysis, we have used the average norm of 1.4. \cite{iyer2015determination} and \cite{Nandi2018} has used this model with mass as a free parameter whereas \cite{Deb2014} has frozen the mass of the central source. We follow the procedure adopted by \cite{iyer2015determination}. We use 3.0 to 150.0 keV {\it RXTE} data for the spectral modelling of outbursts before 2012 and 0.5 to 79.0 keV simultaneous {\it Swift} and {\it NuSTAR} data for spectral modelling of the outbursts after 2012. The results of modelling the energy spectra using the two component accretion flow for the different outbursts of the black hole GX 339-4 are presented in \S  ~\ref{res1}.

\subsection{Method II: QPO evolution} 

Low Frequency QPOs (LFQPOs) are often observed in the power density spectra obtained from fourier transforms of X-ray variability. Of the different types of LFQPOs, C-type QPOs are usually found to be evolving with time. These QPOs are found in the LHS and HIMS of the outburst \citep{Belloni2016}. In the rising phase of the X-ray outburst, the C-type QPO frequencies are found to increase from around 0.1 Hz to about 20 Hz or so \citep{Casella2005}. During the decaying phase of the outburst, QPO frequency evolution exhibits the opposite trend starting at a larger frequency and then gradually decreasing towards lower frequencies. 

The origin of QPOs can be related to oscillating shocks.
We consider the Propagating Oscillatory Shock (POS) model \citep{Chakra2008, nandi2012accretion, radhika2014spectro} that describes the time evolution of C-type QPO frequency. In POS model, QPO is considered to be triggered by an oscillating shock front which propagates towards the black hole during the rising phase of an outburst and moves away from the black hole during the declining phase of the outburst. In this model the QPO frequency is inversely proportional to $r_s^{3/2}$ \citep{CM2000}, where $r_s$ is the instantaneous shock location. 

This model is represented by:

\begin{equation}
\nu _{qpo}=\frac{c}{2\pi R r_g r_s \sqrt{r_s -1}},
\label{eqn:POS1}
\end{equation}

\begin{equation}
r_s = r_{so} \mp \frac{u t + \frac{a t^2}{2}}{r_g},
\label{eqn:POS2}
\end{equation}

where $\nu _{qpo}$ is the QPO frequency, $R$ is the shock compression ratio, $u$ is the initial velocity of the shock front and $a$ its acceleration. $R=\frac{\rho _{+}}{\rho _-}$ is the ratio of density of the post-shock region ($\rho _{+}$) to the density of the pre-shock region ($\rho _{-}$). The instantaneous shock location, $r_s$ and the initial shock location, $r_{so}$ are in units of $r_g$. The negative sign in Equation \ref{eqn:POS2} corresponds to the rising phase and the plus sign corresponds to the decaying phase.

As defined in the previous sub section $r_g$ depends on mass of the black hole which is a parameter of this model. So one can obtain mass of the black hole from POS model fitting as was done by \cite{iyer2015determination, SreeJAA2018}. We apply the POS method on the observed C-type QPOs during the 2002, 2004, 2007 and 2010 outbursts of the source GX 339-4 to estimate the black hole mass as is discussed in \S ~\ref{res2}.

\subsection{Method III: Photon Index-QPO correlation}
It has been observed that the QPO frequency and the photon index of the power-law used to fit the energy spectrum are well correlated. The correlation is modelled using an empirical relation developed by ST07 based on \cite{TF04}. This model is expressed as 
\begin{equation}
\Gamma ( \nu ) = A - BD~ln \left( exp \left( \frac {\nu_{tr} - \nu}{D} \right) + 1\right),
\label{eqn:ST07}
\end{equation}
where $A$ is the index saturation level, $B$ is the initial slope, $D$ determines how fast 
the index goes to saturation and $\nu_{tr}$ is the transition frequency.  Value of parameter $B$ varies in proportion to the mass of the compact object. As shown below, for two sources whose correlation plot has similar saturation the ratio of its mass to $B$ is a constant. 

We model both the target and reference source photon-index versus QPO frequency correlation and estimate the parameters $A$, $B$, $D$ and $\nu_{tr}$. Our intention is to estimate the scaling factor $S_{\nu}$ which is defined as the ratio of mass of the target black hole to the mass of the reference black hole. In equation \ref{eqn:ST07}, if we consider the exponential term to be much larger than unity, we can see that the dimensions of $B$ corresponds to the unit Hz$^{-1}$. Thus, according to equation \ref{eqn:ST07}, $B$ has an inverse relation to QPO frequency. Also, QPO frequencies are inversely related to mass of the black hole (i.e. $ M \propto 1 / \nu $) as is shown in \cite{RemiMc2006} and ST09. This in turn implies that mass of the black hole is directly proportional to the value of parameter B. Thus $\nu _{r} / \nu _{t} = M_t/M_r = S_{\nu} = B_t/B_r$.
%It can be shown that $B$ is a measure of the slope of the correlation plot.% $\left(~ \frac{\Delta \Gamma}{ \Delta \nu}\right)$. 
%{\bf Thus if a reference source with almost similar index saturation levels is used $\frac{\Delta \Gamma _r}{\Delta \Gamma _t} \rightarrow 1,  $ then we have the ratio 
%$ \Delta \nu _{r} /\Delta \nu _{t} = B_t/B_r$, when evaluated at the respective transition frequencies. Here, the subscript $t$ implies target and the subscript $r$ stands for reference. 
%As the lower saturation occurs around zero frequency, we have the ratio of change in frequencies equal to the ratio of frequencies when $\frac{\Delta \Gamma _r}{\Delta \Gamma _t} \rightarrow 1  $. 
%Thus $\nu _{r} / \nu _{t} = S_{\nu} = M_t/M_r = B_t/B_r$.}
%$ B_t/B_r =  \nu _{r} / \nu _{t}=S_{\nu} = M_t/M_r$
 For instance in ST07, mass of Cyg X-1 is estimated to be $8.7\pm 0.8$ from photon index - QPO correlation data of Cyg X-1 spanning over the entire RXTE mission, by scaling with respect to the 2005 outburst (decay phase) of source GRO J1655-40 whose dynamically determined mass is $6.3 \pm 0.5M_{\odot}$ \citep{greene2001optical}. With GRO J1655-40 as the reference source and Cyg X-1 as the target source, we can write  
\begin{equation}
M_{t} = B_{t}\frac{M_{r}}{B_{r}}.
\label{eqn:st07b}
\end{equation} 

The results of this scaling method based on photon index - QPO frequency correlation for the source GX 339-4 are presented in \S ~\ref{res3}.

\section{Results}
\label{results}

The energy spectra from different states of each outburst is modelled with the two component accretion flow which has mass as a free parameter. We also modelled all outburst energy spectra with \textit{pexrav} model following \cite{motta2011low}. The reflection parameter of this model was left free. A multi-colour disc blackbody ({\it diskbb}) and a {\it gaussian} with centroid around 6.4 keV and width 0.8 keV was used whenever necessary. The $N_H$ value of $5 \times 10^{21}~ atoms/cm^2$  used for fitting is obtained from NASA's $N_H$ value calculator \footnote{https://heasarc.gsfc.nasa.gov/cgi-bin/Tools/w3nh/w3nh.pl}. From this modelling, we obtain the photon index of the energy spectra.  Using the temporal and spectral parameters obtained from the above mentioned models fitting, we estimate the mass of the compact object using three indirect methods.    

All the QPO signatures in power spectra are modelled with {\it lorentzians} and the {\it lorentzian} centroid frequency is taken as the QPO frequency. We model the time evolution of the C-type LFQPO frequencies to obtain the BH source mass.% as is presented in \S ~\ref{res2}.

\subsection{Results from Method I}
\label{res1}

\begin{figure}[H]
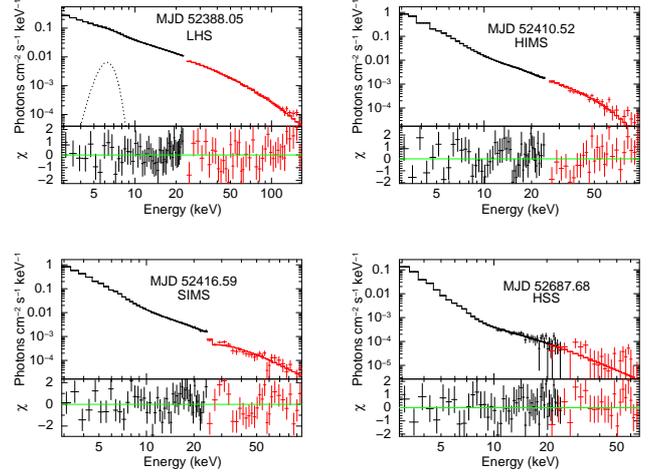


\begin{tabular}{@{}cc@{}}
\includegraphics[scale=0.16,angle=-90]{2002_GX339_LHS_TCAF_modi_1p4.eps} &
\includegraphics[scale=0.16,angle=-90]{2002_GX339_HIMS_TCAF_modi_1p4.eps} \\
\includegraphics[scale=0.16,angle=-90]{2002_GX339_SIMS_TCAF_modi_1p4.eps} &
\includegraphics[scale=0.16,angle=-90]{2002_GX339_HSS_TCAF_modi_1p4.eps}
\end{tabular}
\caption{Spectral fitting with two component flow model (Method I) for 2002 outburst. PCA spectra is shown in black and HEXTE spectra is shown in red. The hard state spectra extends upto 150.0 keV while the soft state spectra extends only upto 60.0 keV for this outburst.}
\label{fig:MethodI_2002}

\end{figure}

\begin{figure}[H]
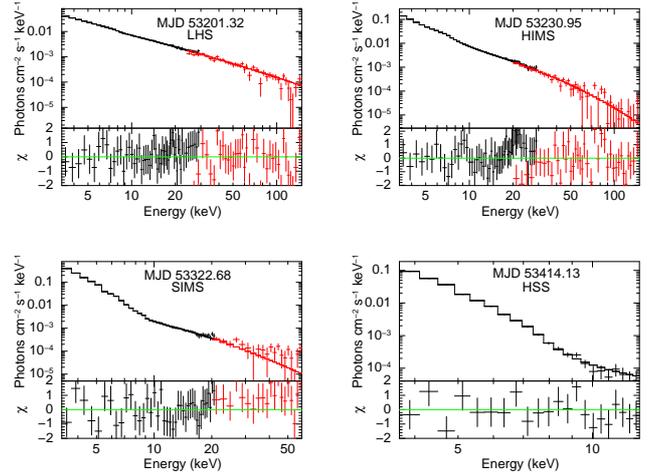


\begin{tabular}{@{}cc@{}}
\includegraphics[scale=0.16,angle=-90]{2004_GX339_LHS_TCAF_1p4.eps}&
\includegraphics[scale=0.16,angle=-90]{2004_GX339_TCAF_HIMS_1p4.eps}\\
\includegraphics[scale=0.16,angle=-90]{2004_GX339_SIMS_TCAF_1p4.eps}&
\includegraphics[scale=0.16,angle=-90]{2004_GX339_HSS_TCAF_1p4.eps}
\end{tabular}
\caption{Spectral fitting with two component flow model (Method I) for the 2004 outburst of GX 339-4. The hard state spectrum extends till 150.0 keV while the soft state spectrum extends only up to 15.0 keV for this outburst.}
\label{fig:MethodI_2004}

\end{figure}

\begin{figure}[H]
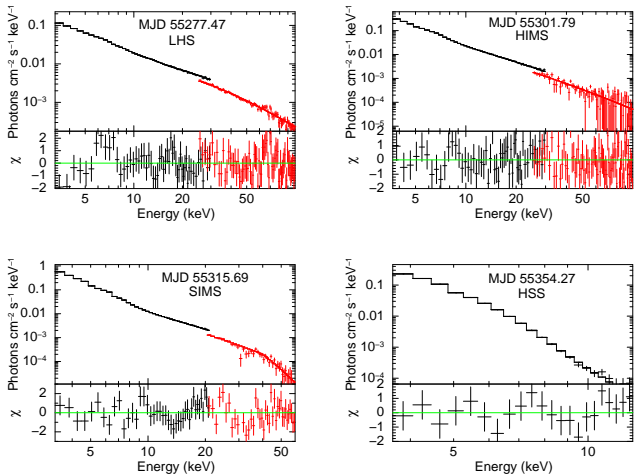


\begin{tabular}{@{}cc@{}}
\includegraphics[scale=0.16,angle=-90]{2010_GX339_LHS_TCAF_1p4.eps}& 
\includegraphics[scale=0.16,angle=-90]{2010_GX339_HIMS_TCAF_1p4.eps}\\
\includegraphics[scale=0.16,angle=-90]{2010_GX339_SIMS_TCAF_1p4.eps}&
\includegraphics[scale=0.16,angle=-90]{2010_GX339_HSS_TCAF_1p4.eps} 
\end{tabular}
\caption{Spectral fitting with two component flow model (Method I) for the 2010 outburst of GX 339-4.}
\label{fig:MethodI_2010}

\end{figure}

\begin{figure}[H]
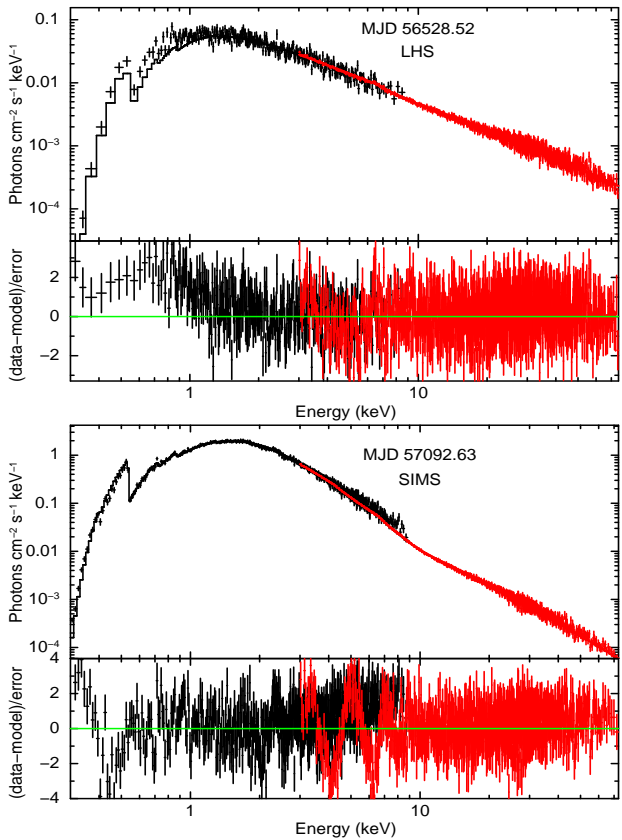

\includegraphics[height=8.5cm,width=5.5cm,angle=-90]{tcaf_gx_2013.eps} 
\includegraphics[height=8.5cm,width=5.5cm,angle=-90]{tcaf_gx_2015.eps} 
\caption{Broadband spectral modelling with Method I for outbursts in 2013 (upper panel) and 2015 (lower panel) showing simultaneous \textit{XRT} (black) and \textit{NuSTAR} (red) spectra. The \textit{XRT} spectral data used, is in the range 0.5 keV to 9.0 keV and the \textit{NuSTAR} data is from 3.0 keV to 79 keV.}
\label{fig:TCAF_2013_15}
\end{figure}

As mentioned in previous section, we have modelled the energy spectrum of the source in all the outbursts observed by RXTE using the two component accretion flow model. Additionally {\it phabs} is used to model the interstellar absorption, {\it smedge} to model the intrinsic absorption, {\it gaussian} to model the iron line and a calibration constant is used to correct for the offset between PCA and HEXTE spectra. We have used both PCA and HEXTE data in the range 3.0 - 150.0 keV for fitting the model. These broadband spectra are shown in Figures \ref{fig:MethodI_2002}, \ref{fig:MethodI_2004} and \ref{fig:MethodI_2010} for 2002, 2004 and 2010 outbursts respectively. The PCA spectra are shown in black and the HEXTE spectra are shown in red. We did not use 2007 outburst data for spectral modelling due to large offset between PCA and HEXTE spectra. Besides this, we also fitted the simultaneous {\it Swift-XRT} and {\it NuSTAR} spectra from the 2013 and 2015 outbursts of the source. This is shown in Figure \ref{fig:TCAF_2013_15}.

We obtain shock location (size of the Compton cloud), halo accretion rate, the disc accretion rate and the mass of the black hole by spectral fitting of each state of the outbursts using this model. We follow the state classification done by \cite{motta2011low} except for the 2013 and 2015 outbursts. For these cases, we have considered the nature of the energy spectra based on photon index values and the contribution of the thermal component to classify the states. 

%Consider the 2002 outburst spectra shown in Figure \ref{fig:MethodI_2002}. 
The energy spectra corresponding to the four states in 2002 outburst of GX 339-4, modelled with the two component accretion flow are shown in Figure \ref{fig:MethodI_2002}. In the LHS shown in top-left panel the spectrum extends from 3.0 keV to 150.0 keV. The shock location corresponding to this spectrum is 274 $r_g$ which implies a large corona around the black hole. The halo accretion rate is 0.83 $\dot{M}_{Edd}$ showing that the contribution of inverse-Comptonized radiation is significant.
This is because the free-falling electrons in the halo can acquire large kinetic energies as they approach the black hole. This triggers thermal Comptonization once kinetic energy has been converted into thermal energy at the shock. The disc accretion rate of 1.70 $\dot{M}_{Edd}$ indicates the presence of a Keplerian disc. As the system transits into the HIMS (top-right panel), the shock location decreases to 92.4 $r_g$, the halo accretion rate changes to 1.12 $\dot{M}_{Edd}$ and the disc accretion rate increases to 4.29 $\dot{M}_{Edd}$. This is evident from the figure where we see comparatively higher photon flux in the 3 to 6 keV range. In the bottom-left panel, we have shown the SIMS spectrum extending from 3 - 100 keV, fitted with the two component flow model. Here the shock location has reached 28 $r_g$ and the halo rate has reduced to 0.48 $\dot{M}_{Edd}$ while the disc rate has become 3.37 $\dot{M}_{Edd}$. Here also the disc is prominent as is evident from the figure where the photon count rate per unit area is high below 6 keV. And finally in the bottom-right panel, we show the fit corresponding to the HSS where the size of the corona has become 25 $r_g$, the halo accretion rate decreased to 0.11 $\dot{M}_{Edd}$ and the disc accretion rate increased to 3.81 $\dot{M}_{Edd}$. This HSS spectrum ranges only from 3 - 60 keV. The mass limits for the BH obtained with this method for the 2002 outburst is 9.04 - 10.61 $M_{\odot}$. 

Similarly, we have modelled the other outbursts observed with RXTE as well and the results are tabulated in Table \ref{tab:massest1a_tab}. It is observed that, generally the disc accretion rate is low in the hard state and then it increases as the states evolve reaching a maximum at the soft state while the halo accretion rate is maximum at the hard states and reduces as the source enters the soft intermediate and high soft states. Also the shock location is found to decrease as the states evolve from LHS to HSS suggesting that the corona becomes compact in the soft states. 

With the \textit{Swift-XRT} and \textit{NuSTAR} data of 2013 and 2015 outburst, we initially applied a phenomenological model consisting of {\it diskbb, powerlaw} and {\it gaussian}. For the 2013 outburst, the {\it powerlaw} is dominant with a photon index of $1.43 \pm 0.01$. The {\it diskbb} component accounts for only 3 \% of the total flux. The source is in LHS during this observation. For the 2015 outburst, the photon index is $2.60 \pm 0.01$ and there is a prominent disc at 0.84 keV implying significant thermal emission. Around 44.8 \% of the total flux is contributed by the thermal component. The source is in its soft intermediate state (SIMS) during this observation. Further, we modelled these two broadband observations (Figure \ref{fig:TCAF_2013_15}) using two component accretion flow. In the observation corresponding to 2013 outburst the shock location is 137 $r_g$, halo rate is 0.54 $\dot{M}_{Edd}$ and has a disc rate of 0.13 $\dot{M}_{Edd}$. During the observation corresponding to 2015 outburst, the shock location is only 15 $r_g$, halo rate is  0.23 $\dot{M}_{Edd}$ and has a significant disc rate of 4.92 $\dot{M}_{Edd}$. The mass estimates at 90 percentage confidence from all the outbursts analysed are given in Table \ref{tab:massest1a_tab}. 

%Apart from the shock location and accretion rates, this model also has mass of the black hole as a parameter. 

%Considering the minimum to maximum mass values obtained from two component flow modelling of all the spectral states together, we get a mass range of 9.04 - 10.61 $M_{\odot}$ for the 2002 outburst, 8.79 - 10.91 $M_{\odot}$ for the 2004 outburst and 8.30 - 10.32 $M_{\odot}$ for the 2010 outburst. Similarly by modelling the broadband spectra available for 2013 and 2015 outbursts we obtain mass values of 9.83 - 9.95 $M_{\odot}$ and 10.13 - 10.21 $M_{\odot}$ respectively. 

%\rotatebox{90}{
\begin{table*}[h]
	\centering
	\caption{Fitted parameters from Two component Flow model. The {\it XRT-NuSTAR} spectra are denoted by $\dagger$. All other observations are from {\it RXTE}.}
	\label{tab:massest1a_tab}
	%\tiny
	\resizebox{\textwidth}{!}{%
	\begin{tabular}{lccccccr} % four columns, alignment for each
		\hline
	    MJD & Outburst & State& $r_s~(r_g)$&$\dot{m}_h~(\dot{M}_{Edd})$&$\dot{m}_d~(\dot{M}_{Edd})$&Mass ($M_{\odot}$)&$\chi ^2/dof$\\
		\hline
	    52388.05 & 2002 & LHS &$274 \pm 15$ &$0.83 \pm 0.01$ &$1.70\pm 0.03$ & $10.49 \pm 0.12$ & 111.42/70\\
		52410.52 & 2002 & HIMS &$92.4 \pm 4$ &$1.12 \pm 0.02$ & $4.29 \pm 0.09$ & $09.59 \pm 0.12$ & 88.79/62\\
		52416.59 & 2002 & SIMS &$28 \pm 1$ &$0.48 \pm 0.002$ &$3.37 \pm 0.07$ & $09.55 \pm 0.07$ & 113.81/62\\
		52687.68 & 2002 & HSS & $25 \pm 1$&$0.11 \pm 0.001$ &$3.81 \pm 0.4$ & $09.49 \pm 0.45$ & 63.94/60\\
		53201.32 & 2004 & LHS & $136 \pm 12$& $0.38 \pm 0.007$& $0.30 \pm 0.02$ & $09.74 \pm 0.32$ & 97.30/90\\
        53230.95 & 2004 & HIMS &$97.5 \pm 1.5$ &$0.27 \pm 0.003$ &$2.27 \pm 0.04$ & $08.89 \pm 0.10$ & 85.89/82\\
		53322.68 & 2004 & SIMS &$10.6 \pm 0.21$ &$0.12 \pm 0.007$ &$6.50 \pm 0.23$ & $10.19 \pm 0.63$ & 66.27/49\\
		53414.13 & 2004 & HSS &$17 \pm 0.5$ &$0.09 \pm 0.001$ &$6.69 \pm 0.25$ & $10.05 \pm 0.86$ & 13.95/10\\
		55277.47 & 2010 & LHS & $453 \pm 34$&$0.30 \pm 0.005$ &$2.9 \pm 0.12$ & $10.16 \pm 0.16$ & 125.72/117\\
		55301.79 & 2010 & HIMS &$43 \pm 1.3$ & $0.29 \pm 0.003$ & $5.9 \pm 0.17$ & $08.97 \pm 0.12$ & 167.97/115\\
		55315.69 & 2010 & SIMS & $23 \pm 0.8$& $0.18\pm 0.002$ &$9.28 \pm 0.38$ & $09.87 \pm 0.27$ & 125.15/79\\
		55354.27 & 2010 & HSS & $5.7 \pm 0.04$ & $0.04 \pm 0.001$ &$7.10 \pm 0.13$ & $08.60 \pm 0.30$ & 14.46/13\\
		56528.52 $\dagger$ & 2013 & LHS &$137 \pm 3$ & $0.54 \pm 0.003$ &$0.13\pm 0.001$ & $09.89 \pm 0.06$ & 1521/1296\\
		57092.63 $\dagger $& 2015 & SIMS &$15 \pm 0.1$ &$0.23 \pm 0.001$ &$4.92 \pm 0.02$ & $10.17 \pm 0.04$ & 2580/1503\\
		\hline
	
     \end{tabular}
     }
     %}
\end{table*}

\subsection{Results from Method II}
\label{res2}

In the POS model, C-Type QPOs which are generally found in LHS and HIMS states of the rising phase of outbursts are considered to be formed by the oscillations of a propagating shock front. So the shock location, shock velocity, shock strength (compression ratio) and mass of the the compact object are the parameters in this model. The model is described by the Equations \ref{eqn:POS1} and \ref{eqn:POS2}. We have used the fitted shock locations from two component flow model for the first day of QPO of each outburst as the initial guess of shock locations for Method II.  

\begin{figure}[H]
\includegraphics[height=7cm,width=1.1\columnwidth]{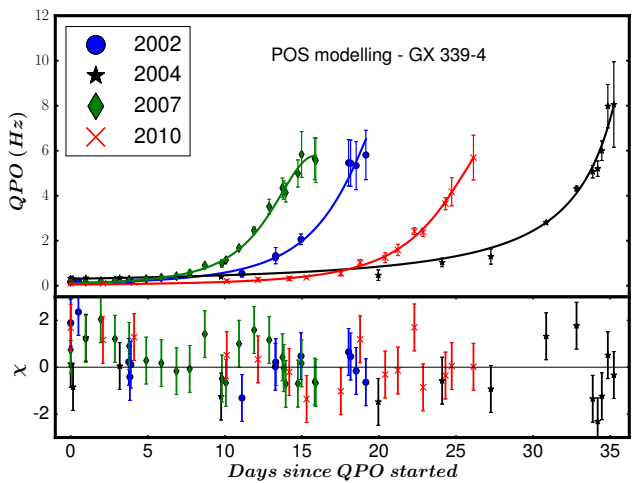} 
\caption{Evolution of QPO frequencies modelled with Method II (POS model) for the rising phase of the outbursts of GX 339-4. The lower panel shows the residuals obtained after POS model fitting.}
\label{fig:cvcd}
\end{figure}

Initially, we have fitted the time evolution of QPO frequencies with constant velocity, but this gives poor fits. For instance the constant velocity based POS model gave a $\chi ^2/dof$ equal to 56/11 for the 2002 outburst QPO evolution. The parameters obtained from this fit are a velocity of $3.7 \pm 0.3~m/s$ and a mass of $8.05 \pm 0.45~M_{\odot}$. Further, we introduced an acceleration parameter to the model and found that it fits the data properly with a $\chi _{red}^2$ close to one.  With the inclusion of the acceleration parameter the fit improved considerably giving a  $\chi ^2/dof$ equal to 12.44/10. Here the initial velocity obtained is $8.5 \pm 0.65~m/s$, the acceleration is $-4.46 \times 10^{-06}~m/s^2$ and a mass range of $9.97 - 10.81~M_{\odot}$. Except for 2004 rising phase data, it is observed that the change in velocity is significant and hence the most general model should include shock front acceleration as a parameter. The POS model fit results for 2002, 2004, 2007 and 2010 outbursts of GX 339-4 are shown in Figure \ref{fig:cvcd}.

For the 2002 outburst as QPO frequency increases from 0.16 Hz to 5.8 Hz the shock location changes from 278 $r_g$ to 18.8 $r_g$ and shock velocity decreases from 8.5 m/s to 6.0 m/s. Similar trend is observed in the rising phases of other outbursts as well. This implies that the shock front slows down as the system transits into SIMS during the rising phase.
%For 2004 outburst as the QPO frequency evolves from 0.30 Hz to 8.05 Hz the shock location changes from 143 $r_g$ to 16.5 $r_g$ and the velocity remains almost constant at 1.2 m/s. The 2007 outburst C-type QPOs evolve from 0.16 Hz to 5.56 Hz while the shock location reduces from 299 $r_g$ to 20.5 $r_g$ and the velocity of the shock front decreases from 12.32 m/s to 8.20 m/s. During the 2010 outburst, the QPO evolves from 0.1 Hz to 5.69 Hz, the shock location varies from 487 $r_g$ to 20.6 $r_g$ and the velocity decreases from 11.72 m/s to 0.89 m/s. 
We also found that during the decay phase of 2007 outburst the QPO frequencies evolved from 4.1 Hz to 0.67 Hz, the corresponding shock location increased from 25.0 $r_g$ to 82.0 $r_g$ and the shock velocity increased from 1.52 m/s to 2.00 m/s. So during the decay phase, the shock front velocity increases as time progresses. In Table \ref{tab:massest_tab2}, we present the values of POS model parameters from the QPO frequency evolution fitting. The overall minimum to maximum mass range obtained from this method is $8.79~-~11.89~M_{\odot}$. 

\begin{table*}[h]
	\centering
	\caption{POS model fitted parameters (Method II)}
	\label{tab:massest_tab2}
	\resizebox{\textwidth}{!}{
	\begin{tabular}{lccccr} % four columns, alignment for each
		\hline
		Outburst &$rs_0(r_g)$ & $v_0(m/s)$ & acceleration ($m/s^2$)&Mass $(M_{\odot})$& $\chi ^2/dof$ \\
		\hline
	    2002 & 278 $\pm$ 4 & 8.50 $\pm$ 0.65  & $-4.46 \times 10^{-06} $ & $9.97-10.81$ & 12.44/10\\
		2004 & 143 $\pm$ 2 & 1.21 $\pm$ 0.09  & $+3.55 \times 10^{-08} $ & $9.97-10.79$ & 17.28/12\\
		2007 & 299 $\pm$ 7 & 12.32 $\pm$ 0.09 & $-9.00 \times 10^{-06} $ & $10.04-10.49$ & 15.87/20\\
		2007decay & 25 $\pm$ 0.6 & 1.52 $\pm$ 0.06  & $+5.00 \times 10^{-07}$ & $9.09-10.99$ & 17.92/15\\
		2010 & 487 $\pm$ 22 & 11.72 $\pm$ 2.15  & $-4.79 \times 10^{-06}$ & $8.79-11.89$ & 14.42/13\\
		\hline
    \end{tabular}
    }
\end{table*}

%\pm 5.84e-07
%\pm 5.38e-08
%-
%-
%\pm 1.06e-06

% It is observed that the QPO evolution for the same source itself is different in duration during different outbursts.

\subsection{Results from Method III}
\label{res3}
As was described earlier the photon index parameter obtained by modelling the energy spectra are correlated with the QPO frequencies of the corresponding power spectra. We fitted this correlation using empirical formulae developed in ST07. We could not apply the  Method III for 2002 outburst data of GX 339-4 as a matching reference outburst is unavailable. So, we used the data from outbursts in 2004, 2007 and 2010 where QPOs are observed for implementing this method. XTE J1550-564 with a dynamically estimated mass of $10.56^{+1.02}_{-0.88}~M_{\odot}$  (\cite{orosz2002dynamical}) and GRO J1655-40 with a dynamically estimated mass of $6.3\pm0.5~M_{\odot}$ \citep{greene2001optical} are the reference sources used.

For the 2004 outburst as the QPO frequency increases from 0.30 Hz to 8.05 Hz the photon index ($\Gamma$) value changes from 1.45 to 2.21. The corresponding reference source used is the 2005 outburst of GRO J1655-40. During the 2007 outburst $\Gamma$ of GX 339-4 evolves from 1.74 to 2.76 as the QPO evolves from its lowest to highest value. The reference source used is the 1998a outburst of XTE J 1550-564. Finally the range of $\Gamma$ for the 2010 outburst is 1.50 to 2.56 which matches the reference outburst 1998b of the source XTE J1550-564. ST09 has classified 1998 outburst of XTE J1550-564  into 1998a and 1998b based on photon index evolution during the outburst.

\begin{figure}[H]
\includegraphics[height=5.5cm,width=8.5cm]{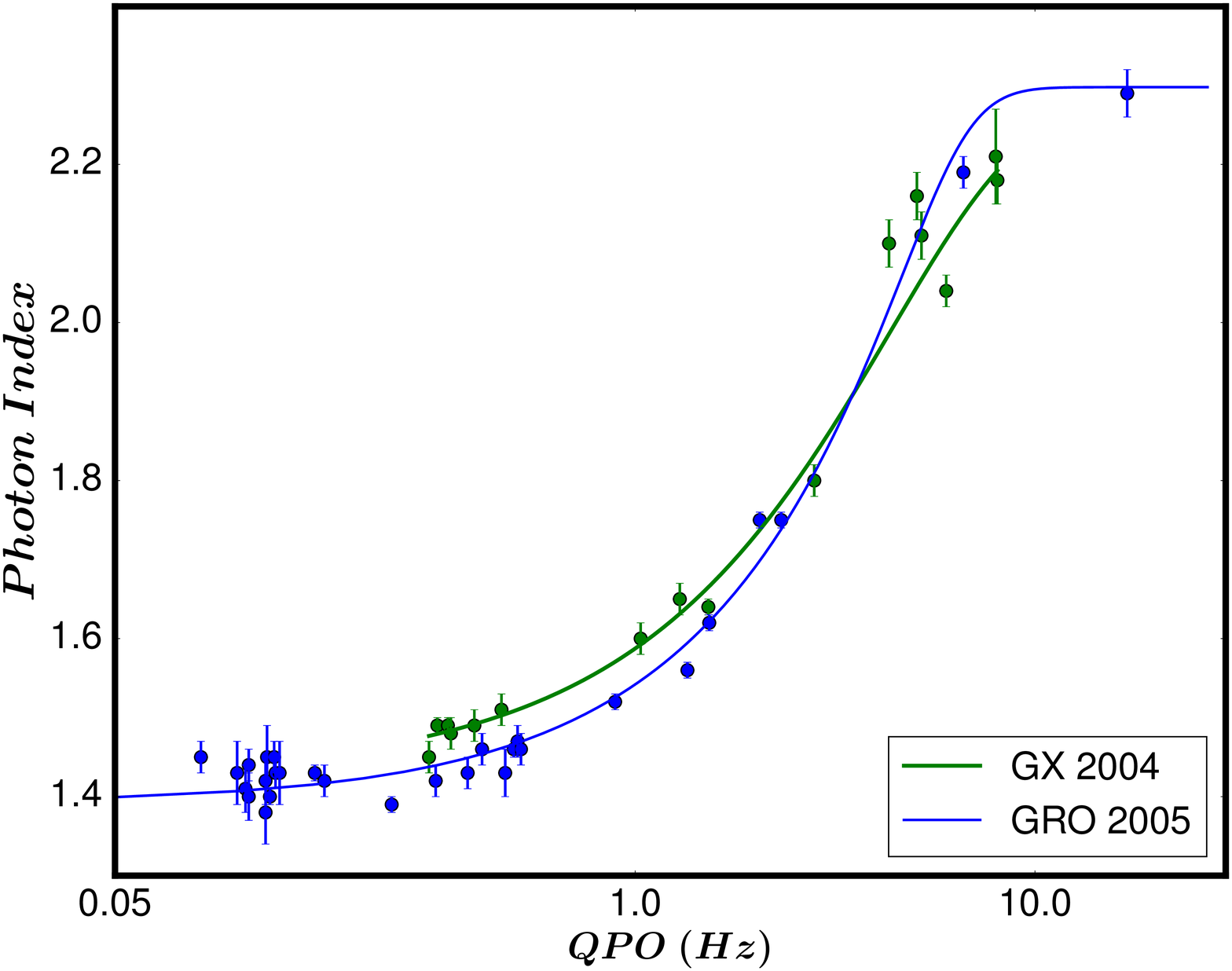}
\includegraphics[height=5.5cm,width=8.5cm]{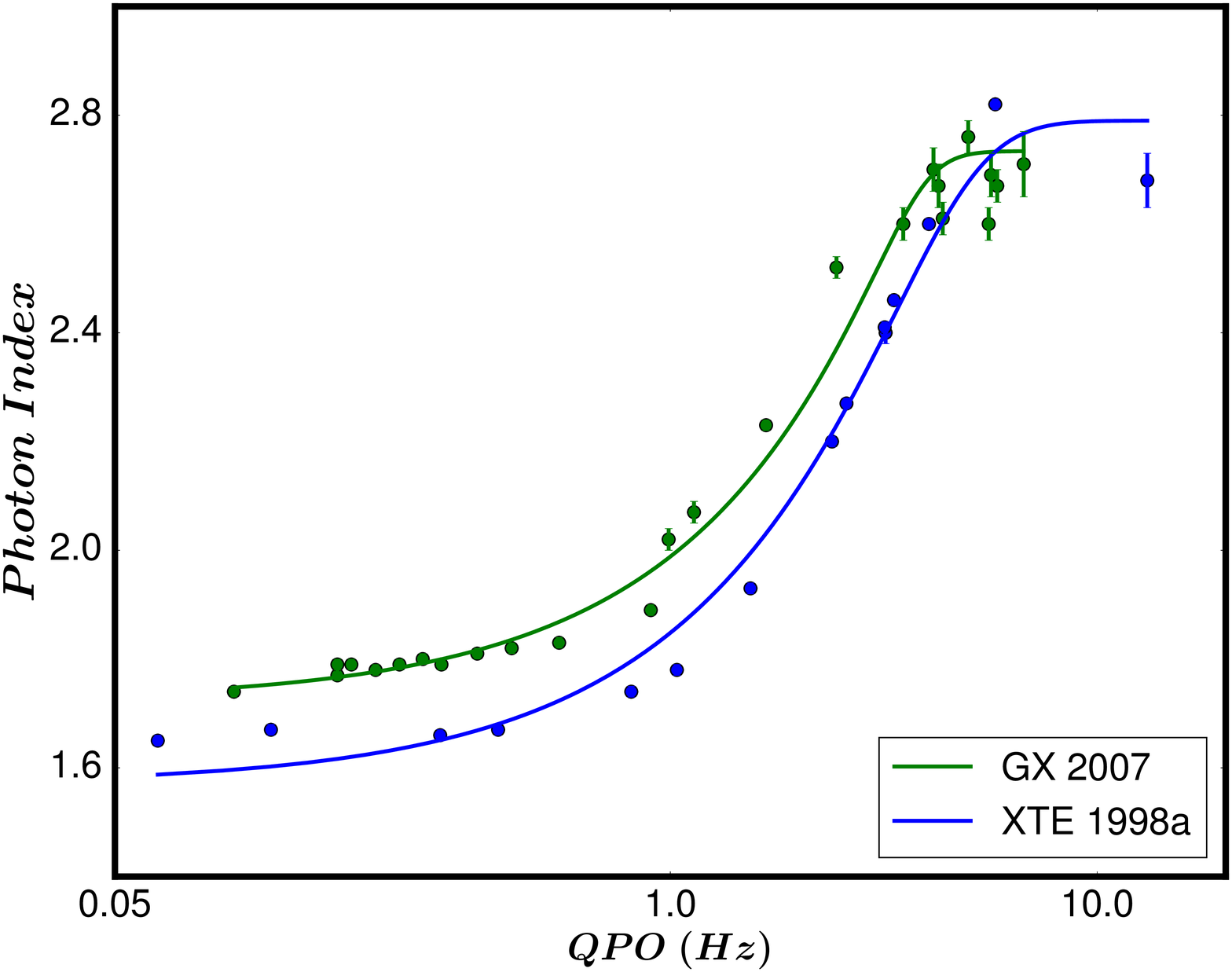}
\includegraphics[height=5.5cm,width=8.5cm]{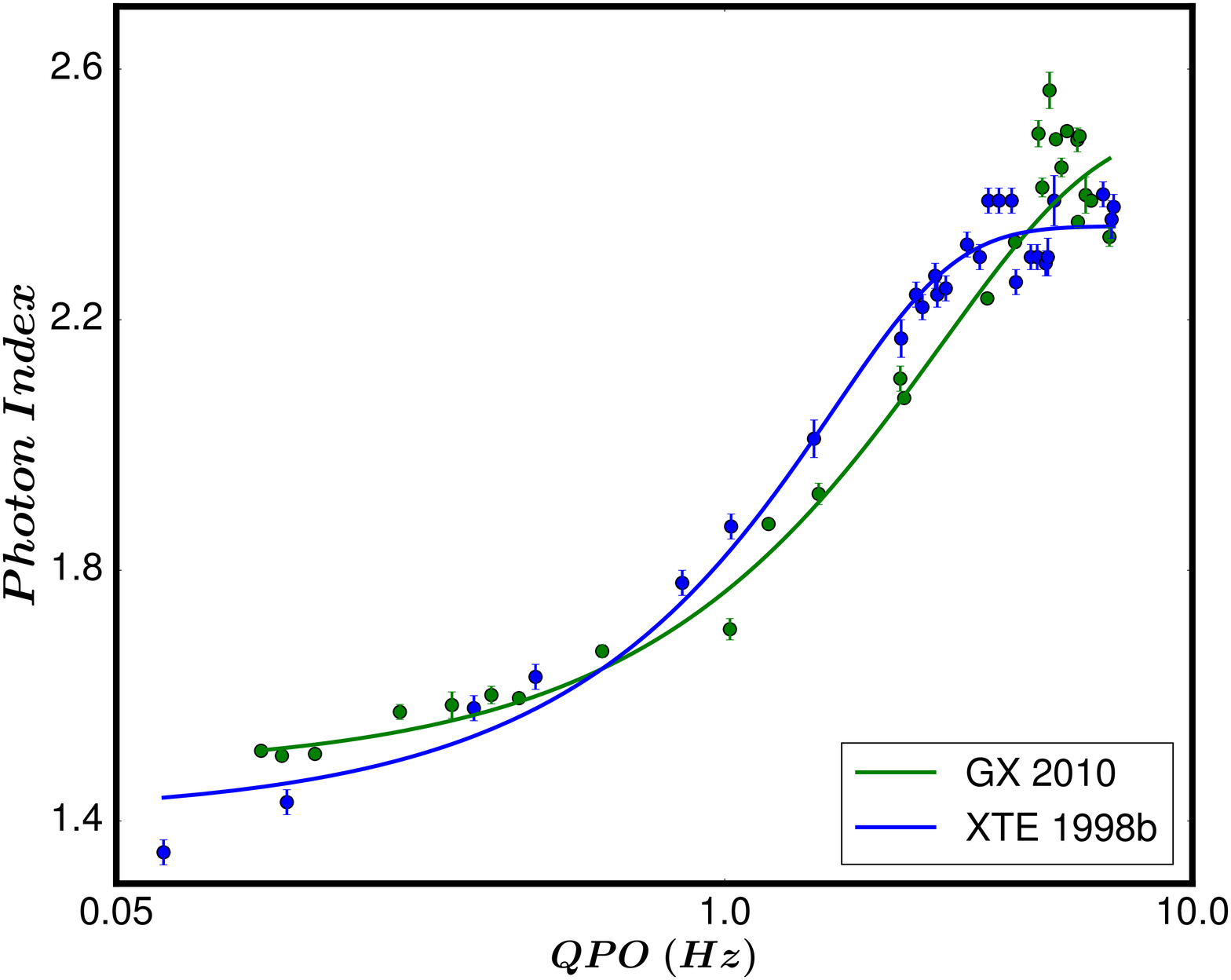}
  
\caption{Fitting of the QPO-photon index correlation based on Method III (ST07). The reference source is shown in blue and the target source in green. The reference sources are chosen such that the photon index saturation levels are similar to that of the target source.}
\label{fig:MethodIII}
\end{figure}

The fits corresponding to Method III are shown in Figure~\ref{fig:MethodIII}. 
%GRO 2005 implies the source name is GRO J1655-40 and the outburst considered is 2005. The label GX 2004 corresponds to year 2004 outburst of the source GX 339-4. Similarly XTE 1998a implies the source is XTE J1550-564 and the outburst considered is 1998a. 
For photon index - QPO frequency correlation, we have used data from same energy ranges for both reference and target. We have used 3 - 30 keV data for the 2004 and 2010 outbursts for both reference and target sources and 3 - 100 keV data for the case of 2007 outburst. This is because the photon index values are significantly different over the two energy ranges and a reference outburst with matching photon index evolution for the 2007 outburst of GX 339-4 was obtained only from the broadband spectral analysis of both the target and reference sources. The results from  Method III presented in this paper are based on the relations given in ST07 (see Equations \ref{eqn:ST07} and \ref{eqn:st07b}).

For the reference source GRO J1655-40 during its 2005 outburst rising phase the parameter $B$ value obtained from the correlation fitting is $0.15_{-0.003}^{+0.01}$. And for 2004 outburst of GX 339-4, we have $B = 0.22_{-0.01}^{+0.02}$. Substituting these values in Equation \ref{eqn:st07b}, we get the mass of GX 339-4 to be in the range $8.50-10.85$ $M_{\odot}$. Similarly we have computed the mass of the target source using data from different outbursts.
In Table \ref{tab:Method3_tab}, we have provided the parameter values obtained by fitting the Gamma ($\Gamma$) - QPO ($\nu$) correlation to different outbursts of GX 339-4 and the reference outbursts used. Thus by using the constant M/B relation in Equation \ref{eqn:st07b}, method III gives an over all mass estimate for GX 339-4 in the range 8.28 to 11.81 $M_{\odot}$. The mass estimates for the different outbursts are given in Table \ref{tab:Method3_tab}.

%The fitting with the empirical formula in ST07 to the reference outburst of XTE J1550-564 during 1998 gives the parameter values $A=2.79,~B=0.285^{+0.031}_{-0.027},~D=1.0$ and $\nu _{tr}=4.26^{+0.14}_{-0.54}~Hz$. The fit results for the 1998b outburst of XTE J1550-564 are $A=2.35 \pm 0.03,~B=0.46^{+0.03}_{-0.02},~D=0.68_{-0.09}^{+0.25}$ and $\nu _{tr}=2.0~Hz$. 

\begin{table*}[h]
	\centering
	\caption{Model fitted parameter values (Method III)}
	\label{tab:Method3_tab}
	%\tiny
	\resizebox{\textwidth}{!}{
	\begin{tabular}{lccccr} % four columns, alignment for each
		\hline
		Source - Outburst & A & B & $\nu_{tr}~(Hz)$ & D & Mass $(M_{\odot})$\\
		\hline
        GX 339 - 2004 &  $2.28_{-0.14}^{+0.34} $ &  $0.22_{-0.01}^{+0.02}$   &  3.0 & $2.72 \pm 0.08$ &$8.50-10.85$\\
		GX 339 - 2007 &  $2.73^{+0.08}_{-0.06} $ &  $0.266_{-0.004}^{+0.035}$ & 3.8 &$0.42 \pm 0.24$ & $8.58-11.81$\\
		%III - 2007 decay &  $2.02 $ &  $0.70 \pm 0.07$ & $0.97 \pm 0.16$ &4.8 &0.06\\
		GX 339 - 2010 &  $2.51_{-0.06}^{+0.09} $ &  $0.40^{+0.02}_{-0.03}$ & 2.0 & $1.81^{+0.69}_{-0.12}$&$8.28-10.49$\\
		\hline 
		Reference - Outburst&&&&&\\
		\hline
		GRO J1655 - 2005  & $2.29 \pm 0.03$       & $0.15_{-0.003}^{+0.01}$ & 6.0 & 1.0 & $6.3 \pm 0.5$\\
		XTE J1550 - 1998a  & $2.79$               &  $0.285^{+0.031}_{-0.027}$& 4.2 & 1.0 & $10.56^{+1.02}_{-0.88}$\\
		XTE J1550 - 1998b  & $2.35 \pm 0.03$      & $0.46^{+0.03}_{-0.02}$   & 2.0 & $0.68_{-0.09}^{+0.25}$ & $10.56^{+1.02}_{-0.88}$\\
		\hline
     \end{tabular}
     }
\end{table*}

\subsection{Consolidated Results}
\begin{figure}[h]
\begin{center}
\includegraphics[width=\columnwidth]{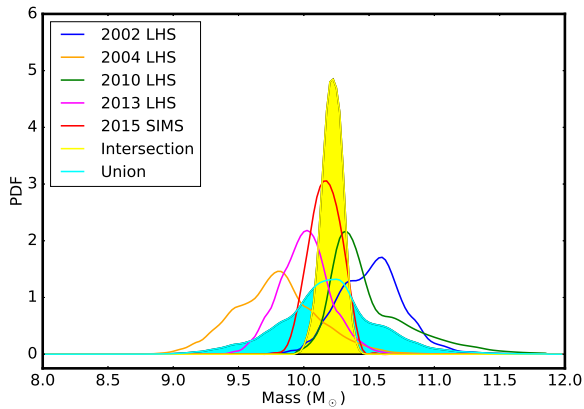}
\end{center}
\caption{The PDFs corresponding to mass estimates from spectral modelling of each outburst is plotted here. Except for 2015 outburst, we have used data from LHS state to obtain the PDFs. The yellow shaded portion is the intersection of individual PDFs and the cyan shaded portion is the union of individual PDFs. See text for more details and mass limits.}
\label{fig:pdf}
\end{figure}

Most of the outbursting black hole sources show sporadic jet activities \citep{Fender2004, Fender2009, Rad2016rxte} and X-ray variabilities \citep{SreeJAA2018,Rad2018} in the intermediate states. So we intend to get a tight constraint on the mass of GX 339-4 using only LHS data. For this we use the results from method I, which is the most robust method among the three that we considered. As broadband LHS data is not available for 2015 outburst, we have used the only available SIMS data for this case. We have converted the outputs of the {\it steppar} command of XSPEC into a probability density function (PDF) following the steps in \cite{Nandi2018}. The same method has been employed in \cite{iyer2015determination} and \cite{Rad2018} to combine different mass estimates. Figure \ref{fig:pdf} shows the PDF corresponding to each outburst
and we have also indicated the union and intersection of these PDFs.  Each PDF is of unit area. The union PDF suggests a mass range of 9.72 - 10.70 $M_{\odot}$ where as the intersection of PDFs gives a tighter constraint of 10.10 - 10.35 $M_{\odot}$. 
%These limits are obtained by computing the area under the PDF such that $\int _{-\infty}^{um}PDF=0.90$ and $\int _{lm}^{\infty}PDF=0.90$ where $um$ is the upper mass limit and $lm$ is the lower mass limit. 
We consider the results from intersection of PDFs, as per the naive Bayes rule for combining independent probabilities \citep{iyer2015determination}. 
%The PDFs from outbursts in 2002 and 2004 peak at slightly
%different masses as compared to other outbursts, but they have finite probability in the intersection range.  

\begin{figure}[h]
\begin{center}
\includegraphics[width=\columnwidth]{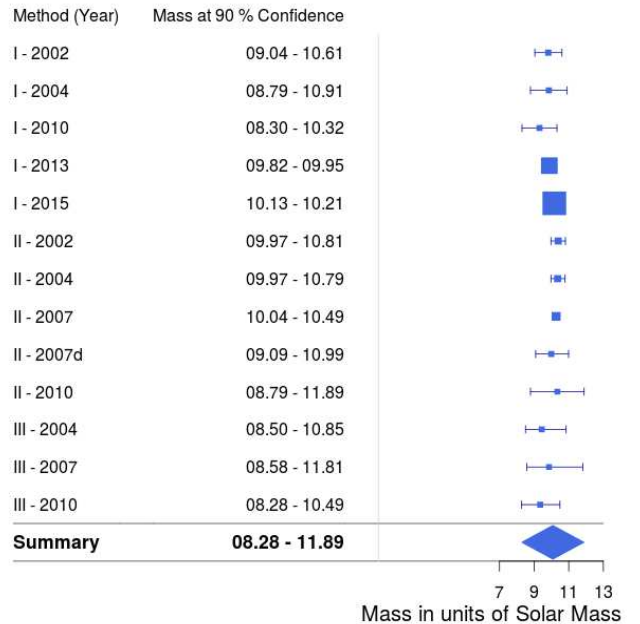}
\end{center} 
\caption{Forest plot of the mass estimates from different methods at 90\% confidence. The size of the square represents the significance of each result. The maximum and minimum values from all three methods are used to plot the final range.}
\label{fig:forest}
\end{figure}

We also include together the results from all the three methods in the Forest plot \footnote{https://cran.r-project.org/web/packages/forestplot/vignettes/forestplot.html} in Figure \ref{fig:forest}. The method and year are shown in the first column, the corresponding mass ranges  at 90\% confidence are shown in the second column and a plot of the mass range is shown in third column. The size of the squares in the plot shows how significant the results are.

As was mentioned earlier, we exclude the 2007 outburst results from Method I due to large offset between PCA and HEXTE spectra. For method II, we have used all rising phase QPO evolution data available from RXTE. We have considered the decay phase data from 2007 as well for POS modelling, but did not include the decay phase data from 2010 outburst \citep{nandi2012accretion} as the QPO evolution is not smooth.  And finally for method III, we have included three outbursts for which we could find matching reference sources. For each individual method, we have considered the minimum and maximum mass values as the range of the mass parameter. Thus the summary measure is given as the minimum to maximum mass values considering all three methods together. So the mass of the black hole in GX 339-4 at a 90\% confidence range is $8.28~-~11.89~ M_{\odot}$.

It should be noted that the PDFs corresponding to Methods II and III will be broad (see \cite{iyer2015determination}). This is also evident from the mass estimates in Tables \ref{tab:massest_tab2} and \ref{tab:Method3_tab} which covers the range obtained from intersection of PDFs from Method I. Hence including PDFs from Methods II and III would not improve the estimate obtained from the joint PDF (intersection) of Method I. The mass estimate from the likelihood analysis (both intersection and union of PDFs) lies within the extreme values obtained from the Forest plot.

\section{Discussion}
\label{disc}
Method I is based on broadband spectral modelling using the two component accretion flow. For this we used {\it RXTE} data in the range 3.0 - 150.0 keV and combined {\it XRT - NuSTAR} data in the range 0.5 - 79.0 keV. From the data of multiple outbursts, the mass of GX 339-4 estimated using this method to be 8.30 - 10.91 $M_{\odot}$. Though the methods we employed are already used by other authors, this is the first time broadband spectral data has been modelled to constrain the mass of the system. The results of spectral modelling from Method I makes it clear that the error bars on mass parameter from {\it RXTE} data analysis (outbursts in 2002 to 2010) is significantly larger than the error bars on parameters obtained with {\it XRT - NuSTAR} data. It can be seen that, the mass range obtained using Method I from 2013 outburst data is $9.82~ - ~9.95~ M_{\odot}$, where as the same method applied to 2015 outburst data gave $10.13 ~- ~10.21~ M_{\odot}$. This is because these two data are obtained when the source was in different states of the outburst. The data systematics will be more in the case of the SIMS as compared to LHS due to presence of other activity like relativistic jet ejections \citep{Fender2004, Fender2009}.
Also, an analytical spectral modelling for optically thin (LHS state) or optically thick (HSS state) conditions produce the required results, but in SIMS the central cloud's (i.e., Compton cloud) Thompson optical depth approaches unity where the analytically calculated spectrum may differ from the expected one (say using monte-carlo simulation) and this may also introduce some amount of uncertainty in the estimation. Besides this one should also consider that the RXTE data starts only from 3 keV onwards, whereas the {\it XRT - NuSTAR} data starts from 0.5 keV onwards and hence we get better constraints
from 2013 and 2015 outburst analysis. 
In this method, we have fixed a constant value of 3 for the compression ratio which is within the theoretical limits of 1 to 4 for hydrodynamic shocks \citep{MC2005}. As compression ratio may vary as the shock location advances and the black hole system transits over different states, it can introduce some uncertainty on the estimated source mass.

In the POS model (Method II), the QPO is considered to originate from the oscillations of a propagating shock front. This model has mass of the black hole as a free parameter. Fitting the QPO frequency evolution using POS model gives us a mass range of $8.79-11.89~M_{\odot}$.
We found that the basic POS model with a constant velocity does not fit the temporal evolution of QPO perfectly. Hence we introduced an acceleration parameter that accounts for the change in velocity of the shock front. We found that the shock front decelerates as the system approaches the soft intermediate state during the rising phase. We also noted that the shock front accelerates as it recedes during the decay phase. During the rising phase of the outburst, particularly in LHS and HIMS, the increase of frequency of QPO with time demands a forward movement of shock front. The location of shock in the accretion flow is determined by the Rankine-Hugoniot conditions which states a balance of both energy and pressure of the flow before and after the shock. The key physical parameters are mass accretion rate and viscosity of the flow. Accretion rate controls the rate of cooling and hence affect the total pressure whereas viscosity decides the transfer of angular momentum. Of course in a magnetized accretion flow, the magnetic pressure also needs to be taken into account.

An increase of accretion rate increases the cooling and shock front moves inward. The same trends are expected with the increases of viscosity as well as strength of magnetic field. In general, the dependency of these effects (rate of cooling, pressure etc.) with the flow parameters are non-linear and hence an accelerated/decelerated movement of the shock front may result along with the evolution of the outburst. A constant speed of shock front is a special case due to internal adjustment of the underlying physical processes. 
%\cite{Sukova2015} has derived the conditions for shock formation and numerically analysed the propagation of shocks in
%the scenario of super massive as well as stellar mass black holes.
\cite{Sukova2015} details the shock formation and numerical analysis of oscillating shock propagation in the scenario of super massive as well as stellar mass black holes. 

In Method II, there is uncertainty in the initial shock location as one is not sure of the exact time at which the QPO has originated. Based on the observations, we consider the time at which the lowest frequency is detected (first instance of detection) as the zero time for modelling QPO frequency evolution in the rising phase. However, this is unlikely to be accurate and hence the mass estimate might also have some uncertainty. Moreover, in this method also, we have considered that the compression ratio is a constant value equal to 3 for all the outbursts. As this need not be the actual scenario, some uncertainty in the estimated mass is possible.

The third method used is the scaling based on photon index - QPO frequency correlation. The over all mass range obtained from this method is $8.28~-~11.81~M_{\odot}$. The disadvantage of this method is the requirement of an appropriate reference source whose photon index saturation levels match with that of the target source. As even the same source does not have similar index saturation levels on different outbursts, it is not possible to find a perfect reference source for this method. Moreover, even if the saturation levels match, the evolution of the photon index is significantly different for target and reference sources.

It must be noted that ST09 has obtained a higher mass estimate with a $\Delta M/M~=~22\%$.
%But in the case of method III, we based our analysis on ST07 instead of ST09. 
ST09 modified equation \ref{eqn:ST07} by introducing a parameter $\beta$ to get, 

\begin{equation}
\Gamma ( \nu ) = A - BD~ln \left( exp \left( \frac {1 - \left(\frac{\nu}{\nu_{tr}}\right)^{\beta} }{D} \right) + 1 \right).
\label{eqn:ST09}
\end{equation}

After getting a good fit using Equation \ref{eqn:ST09} for the reference source correlation, it is 
multiplied by a scaling factor along the frequency axis
in order to get a new scaled curve that matches the photon index-QPO correlation data of the target source. This scaling factor is further
used to estimate the mass of the target source from the mass of the reference source as mass and QPO frequency are inversely related. 
But this geometrical scaling of the fitted curve of the reference pattern 
does not give a perfect match on the target patterns considered. 
Moreover, the scaled curve is quite different from the actual fitted curve of the target pattern.
As a result, here we presented the results of method III based on ST07 rather than ST09.
%Further details about this method are given in ST07.
%However in ST07, parameter $B$ is obtained as the slope of the $\Gamma-\nu$ correlation function given in Equation \ref{eqn:ST07}.
%This implies $\frac{\Delta \Gamma _r}{ \Delta \nu _r}/\frac{\Delta \Gamma _t}{ \Delta \nu _t}=B_r/B_t$.
%For reference and target sources with same index saturation levels the ratio $\frac{\Delta \Gamma _r}{\Delta \Gamma _t}\rightarrow 1$.
%That is $\frac{\Delta \nu _r}{\Delta \nu _t}=\frac{B_t}{B_r}$. This is not equal to the ratio of frequencies and hence to the
%ratio of masses. 

Using the three methods considered in this paper, we have estimated the mass of GX 339-4 to be $8.28~-~11.89~ M_{\odot}$ with a $\Delta M/M$ of 35\%.
The $\Delta M/M$ for \cite{parkermass2016} including both negative and positive error is $31\%$. But \cite{parkermass2016} has used only  data from the very high state, while we have considered multiple outbursts and all different states of the outbursts. Moreover, based on a likelihood analysis we have  obtained the most likely mass of 10.10 - 10.35 $M_{\odot}$ and a broader range of 9.72 - 10.70 $M_{\odot}$ obtained from union of PDFs. Here $\Delta M/M$ is 9.5\%.

\cite{blackcat, Tetarenko2016} provides the mass estimates of more than twenty compact objects in binary systems.
Of this only GRS 1915+105, 4U 1956+350, XTE J1550-564 and 4U 0540-697 
have masses larger than GX 339-4. So in the distribution of dynamical and candidate BH XRBs, 
GX 339-4 is a comparatively higher mass black hole.

%we decided to
%estimate the expression for scaling factor corresponding to Equation \ref{eqn:ST09} as was done in ST07. 

%So rather than shifting the reference correlation curve to match the target pattern, we fitted the the target
%correlation curve also with Equation \ref{eqn:ST09} and obtained the phenomenological fit parameters.
%In this case with the inclusion of $\beta$ parameter, evaluating the slope of the correlation graph gives,
%%\frac{\Delta \nu _r}{\Delta \nu _t}=S_{\nu}= \frac{B_t\beta _t \nu _t^{\beta _t -1} \nu _{r_{tr}}^{\beta _{r}}}{B_r\beta _r \nu _r^{\beta _r -1} \nu _{t_{tr}}^{\beta _{t}}}\frac{\Delta \Gamma _r}{\Delta \Gamma _t} = \frac{B_t\beta _t  \nu _{r_{tr}}}{B_r\beta _r \nu _{t_{tr}}}\frac{\Delta \Gamma _r}{\Delta \Gamma _t}
%
%\begin{equation}
%\frac{ \nu _r}{ \nu _t}=S_{\nu}= \frac{B_t\beta _t  \nu _{r_{tr}}}{B_r\beta _r \nu _{t_{tr}}} ,
%\end{equation}
%%\frac{\Delta \Gamma _r}{\Delta \Gamma _t}
%where the ratio is evaluated at the respective transition frequencies of reference and target. 
%But this method gave inconsistent mass estimates for various outbursts. 

\section{Conclusion} 
\label{conclude}

In this paper, we aimed at determining the mass of the Galactic black hole GX 339-4
using spectral and temporal techniques. We used three different methods to do the mass estimation. Together these methods give a reasonably narrow range of mass for the source. 

The first method we used for mass determination is the two component accretion flow model wherein mass of the compact object is a free parameter. The other parameters in this model are disc and halo accretion rates and shock location. We have tabulated the parameter values obtained by fitting this model to different outburst data in Table \ref{tab:massest1a_tab}. It is usually observed that, the halo rate is dominant in the hard states and the disc rate is dominant in the soft states. Also the shock location which indicates the size of the PSC is larger in the hard states and it decreases as the spectra gets softer. The minimum and maximum mass value obtained from this method over all the outbursts considered is 8.30 $M_{\odot}$ and 10.91 $M_{\odot}$.

Mass determination using the method II, which is based on the POS model also gave us consistent results. Here we studied the temporal evolution of the fundamental QPO frequency as a function of the propagating shock location. As per this model, the C-type LFQPO frequency varies approximately as $r_s^{-3/2}$. 
%We found that the basic POS model with a constant velocity does not fit the temporal evolution of QPO perfectly. Hence we introduced an acceleration parameter that accounts for the change in velocity of the shock front. We found that the shock front decelerates as the system approaches the soft intermediate state during the rising phase. We also noted that the shock front accelerates as it recedes during the decay phase. 
The mass range of the black hole obtained using this method is $8.79-11.89~M_{\odot}$.

The third method is based on the fact that the photon index of the energy spectrum has a correlation with the QPO frequency. As the QPO frequency increases the photon index also increases initially and then saturates. This correlation has been empirically modelled by ST07 with an expression having four variables $A,~B,~D$ and $\nu _{tr}$. We fit the correlation using Equation \ref{eqn:ST07} for a reference source as well as for the target source. The reference source should have similar photon index saturation levels to that of the target source during its outburst. Once the fit parameters for both reference and target source are obtained, we use the scaling relation given in Equation \ref{eqn:st07b} to estimate the mass of the target source. Using this method, the mass estimate of GX 339-4 at 90 \% confidence gives the result $8.28~-~11.81~M_{\odot}$. 
%The disadvantage of this method is the requirement of an appropriate reference source whose photon index saturation levels match with that of the target source. As even the same source does not have similar index saturation levels on different outbursts, it is not possible to find a perfect reference source for this method. Moreover, even if the saturation levels match, the evolution of the photon index is significantly different for target and reference sources. Further details about this method are given in ST07.

The forest plot in Figure \ref{fig:forest} shows all the mass estimates we obtained. We quote the mass estimate based on the three indirect methods to be in the range $8.28~-~11.89~ M_{\odot}$ by considering the overall maximum and minimum values. Previous estimate by \cite{parkermass2016} used only the data from very high state of the source and the results partially overlaps with our estimate. Similarly the mass determined by \cite{Heida17} and \cite{shaposhnikov2009determination} also partially overlaps with our results. Through a likelihood analysis we also infer that the most likely mass for the source is 10.10 - 10.35 $M_{\odot}$. A wider limit of 9.72 - 10.70 $M_{\odot}$ is obtained from the union of PDFs. In method I, we have used data from multiple outbursts and also analysed different states of the BH X-ray binary system in each outburst. In method II, we have used the time evolution of C-type LFQPO frequencies in the LHS and HIMS state of the black hole binary. In method III, we have used the LFQPOs in each outburst appearing in all the states. Thus using a complete spectro-temporal analysis of the outbursts of GX 339-4 we have constrained the mass of the black hole in this system. Among the BHBs whose mass has been estimated so far, GX 339-4 is a comparatively high mass black hole.

\section*{Acknowledgements}
The authors thank both reviewers for their comments and suggestions that helped to improve the quality of the manuscript. This research has made use of data obtained through
the HEASARC web service, provided by the NASA$/$ Goddard Space Flight Center and also made use of data supplied by the UK Swift Science Data Centre at the University of Leicester. We thank GD, SAG; DD, PDMSA and Director, URSC for encouragement and support to carry out this research.

%%%%%%%%%%%%%%%%%%%%%%%%%%%%%%%%%%%%%%%%%%%%%%%%%%%%%%%%%%%%%%%%%%%%%%%%%%%%%
%% Appendices
% \appendix

%\begin{thebibliography}{}
%
%% \bibitem[Names(Year)]{label} or \bibitem[Names(Year)Long names]{label}.
%% (\harvarditem{Name}{Year}{label} is also supported.)
%\bibitem[Elson et al.(1996)]{ESG96}
%Elson, R.A.W., Santiago, B.X., \& Gilmore, G.F. 1996,
%Halo stars, starbursts, and distant globular clusters:
%A survey of unresolved objects in the Hubble Deep Field,
%NewA, 1, 1-16.
%
%\bibitem[Governato et al.(1997)]{Gea97}
%Governato, F., Moore, B., Cen, R., Stadel, J., Lake, G., \& Quinn, T. 1997,
%The Local Group as a test of cosmological models,
%NewA 2, 91-106.
%
%\bibitem[Lamport(1986)]{Lamp86}
%Lamport, L., \LaTeX, {\em A document preparation system},
%2nd edition, Addison-Wesley (Reading, Massachusetts, 1994).
%
%\bibitem[Wettig \& Brown(1996)]{WB96}
%Wettig, T., \& Brown, G.E. 1996,
%The evolution of relativistic binary pulsars,
%NewA, 1, 17-34.
%
%\end{thebibliography}
%\bibliographystyle{plain}
\bibliography{AdSR_gx339}

\begin{thebibliography}{49}
\expandafter\ifx\csname natexlab\endcsname\relax\def\natexlab#1{#1}\fi
\providecommand{\url}[1]{\texttt{#1}}
\providecommand{\href}[2]{#2}
\providecommand{\path}[1]{#1}
\providecommand{\DOIprefix}{doi:}
\providecommand{\ArXivprefix}{arXiv:}
\providecommand{\URLprefix}{URL: }
\providecommand{\Pubmedprefix}{pmid:}
\providecommand{\doi}[1]{\href{http://dx.doi.org/#1}{\path{#1}}}
\providecommand{\Pubmed}[1]{\href{pmid:#1}{\path{#1}}}
\providecommand{\bibinfo}[2]{#2}
\ifx\xfnm\relax \def\xfnm[#1]{\unskip,\space#1}\fi
%Type = Article
\bibitem[{Belloni et~al.(2005)Belloni, Homan, Casella, Van Der~Klis, Nespoli,
  Lewin, Miller and M{\'e}ndez}]{belloni2005evolution}
\bibinfo{author}{Belloni, T.}, \bibinfo{author}{Homan, J.},
  \bibinfo{author}{Casella, P.}, \bibinfo{author}{Van Der~Klis, M.},
  \bibinfo{author}{Nespoli, E.}, \bibinfo{author}{Lewin, W.},
  \bibinfo{author}{Miller, J.}, \bibinfo{author}{M{\'e}ndez, M.},
  \bibinfo{year}{2005}.
\newblock \bibinfo{title}{The evolution of the timing properties of the
  black-hole transient gx 339--4 during its 2002/2003 outburst}.
\newblock \bibinfo{journal}{Astronomy \& Astrophysics} \bibinfo{volume}{440},
  \bibinfo{pages}{207--222}.
%Type = Inproceedings
\bibitem[{{Belloni} and {Motta}(2016)}]{Belloni2016}
\bibinfo{author}{{Belloni}, T.M.}, \bibinfo{author}{{Motta}, S.E.},
  \bibinfo{year}{2016}.
\newblock \bibinfo{title}{{Transient Black Hole Binaries}}, in:
  \bibinfo{editor}{{Bambi}, C.} (Ed.), \bibinfo{booktitle}{Astrophysics of
  Black Holes: From Fundamental Aspects to Latest Developments},
  p.~\bibinfo{pages}{61}.
\newblock \DOIprefix\doi{10.1007/978-3-662-52859-4_2},
  \href{http://arxiv.org/abs/1603.07872}{\tt arXiv:1603.07872}.
%Type = Article
\bibitem[{{Casella} et~al.(2004){Casella}, {Belloni}, {Homan} and
  {Stella}}]{Casella2004}
\bibinfo{author}{{Casella}, P.}, \bibinfo{author}{{Belloni}, T.},
  \bibinfo{author}{{Homan}, J.}, \bibinfo{author}{{Stella}, L.},
  \bibinfo{year}{2004}.
\newblock \bibinfo{title}{{A study of the low-frequency quasi-periodic
  oscillations in the X-ray light curves of the black hole candidate XTE
  J1859+226}}.
\newblock \bibinfo{journal}{Astronomy and Astrophysics} \bibinfo{volume}{426},
  \bibinfo{pages}{587--600}.
\newblock \DOIprefix\doi{10.1051/0004-6361:20041231},
  \href{http://arxiv.org/abs/astro-ph/0407262}{\tt arXiv:astro-ph/0407262}.
%Type = Article
\bibitem[{{Casella} et~al.(2005){Casella}, {Belloni} and
  {Stella}}]{Casella2005}
\bibinfo{author}{{Casella}, P.}, \bibinfo{author}{{Belloni}, T.},
  \bibinfo{author}{{Stella}, L.}, \bibinfo{year}{2005}.
\newblock \bibinfo{title}{{The ABC of Low-Frequency Quasi-periodic Oscillations
  in Black Hole Candidates: Analogies with Z Sources}}.
\newblock \bibinfo{journal}{The Astrophysical Journal} \bibinfo{volume}{629},
  \bibinfo{pages}{403--407}.
\newblock \DOIprefix\doi{10.1086/431174},
  \href{http://arxiv.org/abs/astro-ph/0504318}{\tt arXiv:astro-ph/0504318}.
%Type = Article
\bibitem[{{Chakrabarti} and {Titarchuk}(1995)}]{CT95}
\bibinfo{author}{{Chakrabarti}, S.}, \bibinfo{author}{{Titarchuk}, L.G.},
  \bibinfo{year}{1995}.
\newblock \bibinfo{title}{{Spectral Properties of Accretion Disks around
  Galactic and Extragalactic Black Holes}}.
\newblock \bibinfo{journal}{Astrophysical Journal} \bibinfo{volume}{455},
  \bibinfo{pages}{623}.
\newblock \DOIprefix\doi{10.1086/176610},
  \href{http://arxiv.org/abs/astro-ph/9510005}{\tt arXiv:astro-ph/9510005}.
%Type = Article
\bibitem[{{Chakrabarti} et~al.(2008){Chakrabarti}, {Debnath}, {Nandi} and
  {Pal}}]{Chakra2008}
\bibinfo{author}{{Chakrabarti}, S.K.}, \bibinfo{author}{{Debnath}, D.},
  \bibinfo{author}{{Nandi}, A.}, \bibinfo{author}{{Pal}, P.S.},
  \bibinfo{year}{2008}.
\newblock \bibinfo{title}{{Evolution of the quasi-periodic oscillation
  frequency in GRO J1655-40 - Implications for accretion disk dynamics}}.
\newblock \bibinfo{journal}{Astronomy and Astrophysics} \bibinfo{volume}{489},
  \bibinfo{pages}{L41--L44}.
\newblock \DOIprefix\doi{10.1051/0004-6361:200810136},
  \href{http://arxiv.org/abs/0809.0876}{\tt arXiv:0809.0876}.
%Type = Article
\bibitem[{{Chakrabarti} and {Manickam}(2000)}]{CM2000}
\bibinfo{author}{{Chakrabarti}, S.K.}, \bibinfo{author}{{Manickam}, S.G.},
  \bibinfo{year}{2000}.
\newblock \bibinfo{title}{{Correlation among Quasi-Periodic Oscillation
  Frequencies and Quiescent-State Duration in Black Hole Candidate GRS
  1915+105}}.
\newblock \bibinfo{journal}{Astrophysical Journal Letters}
  \bibinfo{volume}{531}, \bibinfo{pages}{L41--L44}.
\newblock \DOIprefix\doi{10.1086/312512},
  \href{http://arxiv.org/abs/astro-ph/9910012}{\tt arXiv:astro-ph/9910012}.
%Type = Article
\bibitem[{{Corral-Santana} et~al.(2016){Corral-Santana}, {Casares},
  {Mu{\~n}oz-Darias}, {Bauer}, {Mart{\'{\i}}nez-Pais} and {Russell}}]{blackcat}
\bibinfo{author}{{Corral-Santana}, J.M.}, \bibinfo{author}{{Casares}, J.},
  \bibinfo{author}{{Mu{\~n}oz-Darias}, T.}, \bibinfo{author}{{Bauer}, F.E.},
  \bibinfo{author}{{Mart{\'{\i}}nez-Pais}, I.G.}, \bibinfo{author}{{Russell},
  D.M.}, \bibinfo{year}{2016}.
\newblock \bibinfo{title}{{BlackCAT: A catalogue of stellar-mass black holes in
  X-ray transients}}.
\newblock \bibinfo{journal}{Astronomy and Astrophysics} \bibinfo{volume}{587},
  \bibinfo{pages}{A61}.
\newblock \DOIprefix\doi{10.1051/0004-6361/201527130},
  \href{http://arxiv.org/abs/1510.08869}{\tt arXiv:1510.08869}.
%Type = Article
\bibitem[{{Das} et~al.(2001){Das}, {Chattopadhyay} and {Chakrabarti}}]{Das2001}
\bibinfo{author}{{Das}, S.}, \bibinfo{author}{{Chattopadhyay}, I.},
  \bibinfo{author}{{Chakrabarti}, S.K.}, \bibinfo{year}{2001}.
\newblock \bibinfo{title}{{Standing Shocks around Black Holes: An Analytical
  Study}}.
\newblock \bibinfo{journal}{Astrophysical Journal} \bibinfo{volume}{557},
  \bibinfo{pages}{983--989}.
\newblock \DOIprefix\doi{10.1086/321692},
  \href{http://arxiv.org/abs/astro-ph/0107046}{\tt arXiv:astro-ph/0107046}.
%Type = Article
\bibitem[{{Debnath} et~al.(2014){Debnath}, {Chakrabarti} and
  {Mondal}}]{Deb2014}
\bibinfo{author}{{Debnath}, D.}, \bibinfo{author}{{Chakrabarti}, S.K.},
  \bibinfo{author}{{Mondal}, S.}, \bibinfo{year}{2014}.
\newblock \bibinfo{title}{{Implementation of two-component advective flow
  solution in XSPEC}}.
\newblock \bibinfo{journal}{Mon. Not. R. Astron. Soc.} \bibinfo{volume}{440},
  \bibinfo{pages}{L121--L125}.
\newblock \DOIprefix\doi{10.1093/mnrasl/slu024},
  \href{http://arxiv.org/abs/1402.0989}{\tt arXiv:1402.0989}.
%Type = Article
\bibitem[{{Done} and {Kubota}(2006)}]{Done2006}
\bibinfo{author}{{Done}, C.}, \bibinfo{author}{{Kubota}, A.},
  \bibinfo{year}{2006}.
\newblock \bibinfo{title}{{Disc-corona energetics in the very high state of
  Galactic black holes}}.
\newblock \bibinfo{journal}{Mon. Not. R. Astron. Soc.} \bibinfo{volume}{371},
  \bibinfo{pages}{1216--1230}.
\newblock \DOIprefix\doi{10.1111/j.1365-2966.2006.10737.x},
  \href{http://arxiv.org/abs/astro-ph/0511030}{\tt arXiv:astro-ph/0511030}.
%Type = Article
\bibitem[{{Evans} et~al.(2009){Evans}, {Beardmore}, {Page}, {Osborne},
  {O'Brien}, {Willingale}, {Starling}, {Burrows}, {Godet}, {Vetere}, {Racusin},
  {Goad}, {Wiersema}, {Angelini}, {Capalbi}, {Chincarini}, {Gehrels}, {Kennea},
  {Margutti}, {Morris}, {Mountford}, {Pagani}, {Perri}, {Romano} and
  {Tanvir}}]{Evans2009}
\bibinfo{author}{{Evans}, P.A.}, \bibinfo{author}{{Beardmore}, A.P.},
  \bibinfo{author}{{Page}, K.L.}, \bibinfo{author}{{Osborne}, J.P.},
  \bibinfo{author}{{O'Brien}, P.T.}, \bibinfo{author}{{Willingale}, R.},
  \bibinfo{author}{{Starling}, R.L.C.}, \bibinfo{author}{{Burrows}, D.N.},
  \bibinfo{author}{{Godet}, O.}, \bibinfo{author}{{Vetere}, L.},
  \bibinfo{author}{{Racusin}, J.}, \bibinfo{author}{{Goad}, M.R.},
  \bibinfo{author}{{Wiersema}, K.}, \bibinfo{author}{{Angelini}, L.},
  \bibinfo{author}{{Capalbi}, M.}, \bibinfo{author}{{Chincarini}, G.},
  \bibinfo{author}{{Gehrels}, N.}, \bibinfo{author}{{Kennea}, J.A.},
  \bibinfo{author}{{Margutti}, R.}, \bibinfo{author}{{Morris}, D.C.},
  \bibinfo{author}{{Mountford}, C.J.}, \bibinfo{author}{{Pagani}, C.},
  \bibinfo{author}{{Perri}, M.}, \bibinfo{author}{{Romano}, P.},
  \bibinfo{author}{{Tanvir}, N.}, \bibinfo{year}{2009}.
\newblock \bibinfo{title}{{Methods and results of an automatic analysis of a
  complete sample of Swift-XRT observations of GRBs}}.
\newblock \bibinfo{journal}{Mon. Not. R. Astron. Soc.} \bibinfo{volume}{397},
  \bibinfo{pages}{1177--1201}.
\newblock \DOIprefix\doi{10.1111/j.1365-2966.2009.14913.x},
  \href{http://arxiv.org/abs/0812.3662}{\tt arXiv:0812.3662}.
%Type = Article
\bibitem[{{Fender} et~al.(2004){Fender}, {Belloni} and {Gallo}}]{Fender2004}
\bibinfo{author}{{Fender}, R.P.}, \bibinfo{author}{{Belloni}, T.M.},
  \bibinfo{author}{{Gallo}, E.}, \bibinfo{year}{2004}.
\newblock \bibinfo{title}{{Towards a unified model for black hole X-ray binary
  jets}}.
\newblock \bibinfo{journal}{Mon. Not. R. Astron. Soc.} \bibinfo{volume}{355},
  \bibinfo{pages}{1105--1118}.
\newblock \DOIprefix\doi{10.1111/j.1365-2966.2004.08384.x},
  \href{http://arxiv.org/abs/astro-ph/0409360}{\tt arXiv:astro-ph/0409360}.
%Type = Article
\bibitem[{{Fender} et~al.(2009){Fender}, {Homan} and {Belloni}}]{Fender2009}
\bibinfo{author}{{Fender}, R.P.}, \bibinfo{author}{{Homan}, J.},
  \bibinfo{author}{{Belloni}, T.M.}, \bibinfo{year}{2009}.
\newblock \bibinfo{title}{{Jets from black hole X-ray binaries: testing,
  refining and extending empirical models for the coupling to X-rays}}.
\newblock \bibinfo{journal}{Mon. Not. R. Astron. Soc.} \bibinfo{volume}{396},
  \bibinfo{pages}{1370--1382}.
\newblock \DOIprefix\doi{10.1111/j.1365-2966.2009.14841.x},
  \href{http://arxiv.org/abs/0903.5166}{\tt arXiv:0903.5166}.
%Type = Article
\bibitem[{{Giri} and {Chakrabarti}(2013)}]{GiriSKC2013}
\bibinfo{author}{{Giri}, K.}, \bibinfo{author}{{Chakrabarti}, S.K.},
  \bibinfo{year}{2013}.
\newblock \bibinfo{title}{{Hydrodynamic simulation of two-component advective
  flows around black holes}}.
\newblock \bibinfo{journal}{Mon. Not. R. Astron. Soc.} \bibinfo{volume}{430},
  \bibinfo{pages}{2836--2843}.
\newblock \DOIprefix\doi{10.1093/mnras/stt087},
  \href{http://arxiv.org/abs/1212.6493}{\tt arXiv:1212.6493}.
%Type = Article
\bibitem[{Grebenev et~al.(1991)Grebenev, Sunyaev, Pavlinsky and
  Dekhanov}]{grebenev1991detection}
\bibinfo{author}{Grebenev, S.}, \bibinfo{author}{Sunyaev, R.},
  \bibinfo{author}{Pavlinsky, M.}, \bibinfo{author}{Dekhanov, I.},
  \bibinfo{year}{1991}.
\newblock \bibinfo{title}{Detection of quasi-periodic oscillations of
  x-radiation from the black hole candidate gx339-4}.
\newblock \bibinfo{journal}{Pisma v Astronomicheskii Zhurnal}
  \bibinfo{volume}{17}, \bibinfo{pages}{985--990}.
%Type = Article
\bibitem[{Greene et~al.(2001)Greene, Bailyn and Orosz}]{greene2001optical}
\bibinfo{author}{Greene, J.}, \bibinfo{author}{Bailyn, C.D.},
  \bibinfo{author}{Orosz, J.A.}, \bibinfo{year}{2001}.
\newblock \bibinfo{title}{Optical and infrared photometry of the microquasar
  gro j1655--40 in quiescence}.
\newblock \bibinfo{journal}{The Astrophysical Journal} \bibinfo{volume}{554},
  \bibinfo{pages}{1290}.
%Type = Article
\bibitem[{{Heida} et~al.(2017){Heida}, {Jonker}, {Torres} and
  {Chiavassa}}]{Heida17}
\bibinfo{author}{{Heida}, M.}, \bibinfo{author}{{Jonker}, P.G.},
  \bibinfo{author}{{Torres}, M.A.P.}, \bibinfo{author}{{Chiavassa}, A.},
  \bibinfo{year}{2017}.
\newblock \bibinfo{title}{{The Mass Function of GX 339-4 from Spectroscopic
  Observations of Its Donor Star}}.
\newblock \bibinfo{journal}{Astrophysical Journal} \bibinfo{volume}{846},
  \bibinfo{pages}{132}.
\newblock \DOIprefix\doi{10.3847/ 1538-4357/ aa85df},
  \href{http://arxiv.org/abs/1708.04667}{\tt arXiv:1708.04667}.
%Type = Article
\bibitem[{{Homan} and {Belloni}(2005)}]{HoBel2005}
\bibinfo{author}{{Homan}, J.}, \bibinfo{author}{{Belloni}, T.},
  \bibinfo{year}{2005}.
\newblock \bibinfo{title}{{The Evolution of Black Hole States}}.
\newblock \bibinfo{journal}{Astrophysics and Space Science}
  \bibinfo{volume}{300}, \bibinfo{pages}{107--117}.
\newblock \DOIprefix\doi{10.1007/s10509-005-1197-4},
  \href{http://arxiv.org/abs/astro-ph/0412597}{\tt arXiv:astro-ph/0412597}.
%Type = Article
\bibitem[{{Homan} et~al.(2001){Homan}, {Wijnands}, {van der Klis}, {Belloni},
  {van Paradijs}, {Klein-Wolt}, {Fender} and {M{\'e}ndez}}]{2001Homan}
\bibinfo{author}{{Homan}, J.}, \bibinfo{author}{{Wijnands}, R.},
  \bibinfo{author}{{van der Klis}, M.}, \bibinfo{author}{{Belloni}, T.},
  \bibinfo{author}{{van Paradijs}, J.}, \bibinfo{author}{{Klein-Wolt}, M.},
  \bibinfo{author}{{Fender}, R.}, \bibinfo{author}{{M{\'e}ndez}, M.},
  \bibinfo{year}{2001}.
\newblock \bibinfo{title}{{Correlated X-Ray Spectral and Timing Behavior of the
  Black Hole Candidate XTE J1550-564: A New Interpretation of Black Hole
  States}}.
\newblock \bibinfo{journal}{Astrophys. J. Suppl. Ser.} \bibinfo{volume}{132},
  \bibinfo{pages}{377--402}.
\newblock \DOIprefix\doi{10.1086/318954},
  \href{http://arxiv.org/abs/astro-ph/0001163}{\tt arXiv:astro-ph/0001163}.
%Type = Article
\bibitem[{Hynes et~al.(2003)Hynes, Steeghs, Casares, Charles and
  O'Brien}]{hynes2003dynamical}
\bibinfo{author}{Hynes, R.I.}, \bibinfo{author}{Steeghs, D.},
  \bibinfo{author}{Casares, J.}, \bibinfo{author}{Charles, P.},
  \bibinfo{author}{O'Brien, K.}, \bibinfo{year}{2003}.
\newblock \bibinfo{title}{Dynamical evidence for a black hole in gx 339--4}.
\newblock \bibinfo{journal}{The Astrophysical Journal Letters}
  \bibinfo{volume}{583}, \bibinfo{pages}{L95}.
%Type = Article
\bibitem[{{Ingram} and {Done}(2011)}]{Ingram2011}
\bibinfo{author}{{Ingram}, A.}, \bibinfo{author}{{Done}, C.},
  \bibinfo{year}{2011}.
\newblock \bibinfo{title}{{A physical model for the continuum variability and
  quasi-periodic oscillation in accreting black holes}}.
\newblock \bibinfo{journal}{Mon. Not. R. Astron. Soc.} \bibinfo{volume}{415},
  \bibinfo{pages}{2323--2335}.
\newblock \DOIprefix\doi{10.1111/j.1365-2966.2011.18860.x},
  \href{http://arxiv.org/abs/1101.2336}{\tt arXiv:1101.2336}.
%Type = Article
\bibitem[{{Iyer} et~al.(2015){Iyer}, {Nandi} and
  {Mandal}}]{iyer2015determination}
\bibinfo{author}{{Iyer}, N.}, \bibinfo{author}{{Nandi}, A.},
  \bibinfo{author}{{Mandal}, S.}, \bibinfo{year}{2015}.
\newblock \bibinfo{title}{{Determination of the Mass of IGR J17091-3624 from
  ``Spectro-temporal'' Variations during the Onset Phase of the 2011
  Outburst}}.
\newblock \bibinfo{journal}{Astrophysical Journal} \bibinfo{volume}{807},
  \bibinfo{pages}{108}.
\newblock \DOIprefix\doi{10.1088/0004-637X/807/1/108},
  \href{http://arxiv.org/abs/1505.02529}{\tt arXiv:1505.02529}.
%Type = Article
\bibitem[{{Maccarone} and {Coppi}(2003)}]{Macncop2003}
\bibinfo{author}{{Maccarone}, T.J.}, \bibinfo{author}{{Coppi}, P.S.},
  \bibinfo{year}{2003}.
\newblock \bibinfo{title}{{Hysteresis in the light curves of soft X-ray
  transients}}.
\newblock \bibinfo{journal}{Mon. Not. R. Astron. Soc.} \bibinfo{volume}{338},
  \bibinfo{pages}{189--196}.
\newblock \DOIprefix\doi{10.1046/j.1365-8711.2003.06040.x},
  \href{http://arxiv.org/abs/astro-ph/0209116}{\tt arXiv:astro-ph/0209116}.
%Type = Article
\bibitem[{{Mandal} and {Chakrabarti}(2005)}]{MC2005}
\bibinfo{author}{{Mandal}, S.}, \bibinfo{author}{{Chakrabarti}, S.K.},
  \bibinfo{year}{2005}.
\newblock \bibinfo{title}{{Accretion shock signatures in the spectrum of
  two-temperature advective flows around black holes}}.
\newblock \bibinfo{journal}{Astronomy and Astrophysics} \bibinfo{volume}{434},
  \bibinfo{pages}{839--848}.
\newblock \DOIprefix\doi{10.1051/0004-6361:20041235}.
%Type = Article
\bibitem[{Motta et~al.(2011)Motta, Munoz-Darias, Casella, Belloni and
  Homan}]{motta2011low}
\bibinfo{author}{Motta, S.}, \bibinfo{author}{Munoz-Darias, T.},
  \bibinfo{author}{Casella, P.}, \bibinfo{author}{Belloni, T.},
  \bibinfo{author}{Homan, J.}, \bibinfo{year}{2011}.
\newblock \bibinfo{title}{Low-frequency oscillations in black holes: a
  spectral-timing approach to the case of gx 339-4}.
\newblock \bibinfo{journal}{Mon. Not. R. Astron. Soc.} \bibinfo{volume}{418},
  \bibinfo{pages}{2292--2307}.
%Type = Article
\bibitem[{Munoz-Darias et~al.(2008)Munoz-Darias, Casares and
  Mart{\'\i}nez-Pais}]{munoz2008masses}
\bibinfo{author}{Munoz-Darias, T.}, \bibinfo{author}{Casares, J.},
  \bibinfo{author}{Mart{\'\i}nez-Pais, I.}, \bibinfo{year}{2008}.
\newblock \bibinfo{title}{On the masses and evolutionary status of the black
  hole binary gx 339-4: a twin system of xte j1550-564?}
\newblock \bibinfo{journal}{Monthly Notices of the Royal Astronomical Society}
  \bibinfo{volume}{385}, \bibinfo{pages}{2205--2209}.
%Type = Article
\bibitem[{Nandi et~al.(2012)Nandi, Debnath, Mandal and
  Chakrabarti}]{nandi2012accretion}
\bibinfo{author}{Nandi, A.}, \bibinfo{author}{Debnath, D.},
  \bibinfo{author}{Mandal, S.}, \bibinfo{author}{Chakrabarti, S.K.},
  \bibinfo{year}{2012}.
\newblock \bibinfo{title}{Accretion flow dynamics during the evolution of
  timing and spectral properties of gx 339-4 during its 2010--11 outburst}.
\newblock \bibinfo{journal}{Astronomy \& Astrophysics} \bibinfo{volume}{542},
  \bibinfo{pages}{A56}.
%Type = Article
\bibitem[{{Nandi} et~al.(2018){Nandi}, {Mandal}, {Sreehari}, {Radhika}, {Das},
  {Chattopadhyay}, {Iyer}, {Agrawal} and {Aktar}}]{Nandi2018}
\bibinfo{author}{{Nandi}, A.}, \bibinfo{author}{{Mandal}, S.},
  \bibinfo{author}{{Sreehari}, H.}, \bibinfo{author}{{Radhika}, D.},
  \bibinfo{author}{{Das}, S.}, \bibinfo{author}{{Chattopadhyay}, I.},
  \bibinfo{author}{{Iyer}, N.}, \bibinfo{author}{{Agrawal}, V.K.},
  \bibinfo{author}{{Aktar}, R.}, \bibinfo{year}{2018}.
\newblock \bibinfo{title}{{Accretion flow dynamics during 1999 outburst of XTE
  J1859+226-modeling of broadband spectra and constraining the source mass}}.
\newblock \bibinfo{journal}{Astrophysics \& Space Science}
  \bibinfo{volume}{363}, \bibinfo{pages}{90}.
\newblock \DOIprefix\doi{10.1007/s10509-018-3314-1},
  \href{http://arxiv.org/abs/1803.08638}{\tt arXiv:1803.08638}.
%Type = Article
\bibitem[{Orosz et~al.(2002)Orosz, Groot, van~der Klis, McClintock, Garcia,
  Zhao, Jain, Bailyn and Remillard}]{orosz2002dynamical}
\bibinfo{author}{Orosz, J.A.}, \bibinfo{author}{Groot, P.J.},
  \bibinfo{author}{van~der Klis, M.}, \bibinfo{author}{McClintock, J.E.},
  \bibinfo{author}{Garcia, M.R.}, \bibinfo{author}{Zhao, P.},
  \bibinfo{author}{Jain, R.K.}, \bibinfo{author}{Bailyn, C.D.},
  \bibinfo{author}{Remillard, R.A.}, \bibinfo{year}{2002}.
\newblock \bibinfo{title}{Dynamical evidence for a black hole in the
  microquasar xte j1550--564}.
\newblock \bibinfo{journal}{The Astrophysical Journal} \bibinfo{volume}{568},
  \bibinfo{pages}{845}.
%Type = Article
\bibitem[{{Parker} et~al.(2016){Parker}, {Tomsick}, {Kennea}, {Miller},
  {Harrison}, {Barret}, {Boggs}, {Christensen}, {Craig}, {Fabian}, {F{\"u}rst},
  {Grinberg}, {Hailey}, {Romano}, {Stern}, {Walton} and
  {Zhang}}]{parkermass2016}
\bibinfo{author}{{Parker}, M.L.}, \bibinfo{author}{{Tomsick}, J.A.},
  \bibinfo{author}{{Kennea}, J.A.}, \bibinfo{author}{{Miller}, J.M.},
  \bibinfo{author}{{Harrison}, F.A.}, \bibinfo{author}{{Barret}, D.},
  \bibinfo{author}{{Boggs}, S.E.}, \bibinfo{author}{{Christensen}, F.E.},
  \bibinfo{author}{{Craig}, W.W.}, \bibinfo{author}{{Fabian}, A.C.},
  \bibinfo{author}{{F{\"u}rst}, F.}, \bibinfo{author}{{Grinberg}, V.},
  \bibinfo{author}{{Hailey}, C.J.}, \bibinfo{author}{{Romano}, P.},
  \bibinfo{author}{{Stern}, D.}, \bibinfo{author}{{Walton}, D.J.},
  \bibinfo{author}{{Zhang}, W.W.}, \bibinfo{year}{2016}.
\newblock \bibinfo{title}{{NuSTAR and Swift Observations of the Very High State
  in GX 339-4: Weighing the Black Hole with X-Rays}}.
\newblock \bibinfo{journal}{Astrophys. J. Lett.} \bibinfo{volume}{821},
  \bibinfo{pages}{L6}.
\newblock \DOIprefix\doi{10.3847/2041-8205/821/1/L6},
  \href{http://arxiv.org/abs/1603.03777}{\tt arXiv:1603.03777}.
%Type = Article
\bibitem[{Radhika and Nandi(2014)}]{radhika2014spectro}
\bibinfo{author}{Radhika, D.}, \bibinfo{author}{Nandi, A.},
  \bibinfo{year}{2014}.
\newblock \bibinfo{title}{‘spectro-temporal’characteristics and disk-jet
  connection of the outbursting black hole source xte j1859+ 226}.
\newblock \bibinfo{journal}{Advances in Space Research} \bibinfo{volume}{54},
  \bibinfo{pages}{1678--1697}.
%Type = Article
\bibitem[{{Radhika} et~al.(2016){Radhika}, {Nandi}, {Agrawal} and
  {Seetha}}]{Rad2016rxte}
\bibinfo{author}{{Radhika}, D.}, \bibinfo{author}{{Nandi}, A.},
  \bibinfo{author}{{Agrawal}, V.K.}, \bibinfo{author}{{Seetha}, S.},
  \bibinfo{year}{2016}.
\newblock \bibinfo{title}{{`Spectro-temporal' variabilities and possible
  physical mechanism for jet ejections}}.
\newblock \bibinfo{journal}{Mon. Not. R. Astron. Soc.} \bibinfo{volume}{460},
  \bibinfo{pages}{4403--4416}.
\newblock \DOIprefix\doi{10.1093/mnras/stw1239},
  \href{http://arxiv.org/abs/1605.08351}{\tt arXiv:1605.08351}.
%Type = Article
\bibitem[{{Radhika} et~al.(2018){Radhika}, {Sreehari}, {Nandi}, {Iyer} and
  {Mandal}}]{Rad2018}
\bibinfo{author}{{Radhika}, D.}, \bibinfo{author}{{Sreehari}, H.},
  \bibinfo{author}{{Nandi}, A.}, \bibinfo{author}{{Iyer}, N.},
  \bibinfo{author}{{Mandal}, S.}, \bibinfo{year}{2018}.
\newblock \bibinfo{title}{{Broad-band spectral evolution and temporal
  variability of IGR J17091-3624 during its 2016 outburst: SWIFT and NuSTAR
  results}}.
\newblock \bibinfo{journal}{Astrophysics \& Space Science}
  \bibinfo{volume}{363}, \bibinfo{pages}{189}.
\newblock \DOIprefix\doi{10.1007/s10509-018-3411-1},
  \href{http://arxiv.org/abs/1808.05556}{\tt arXiv:1808.05556}.
%Type = Article
\bibitem[{{Remillard} and {McClintock}(2006)}]{RemiMc2006}
\bibinfo{author}{{Remillard}, R.A.}, \bibinfo{author}{{McClintock}, J.E.},
  \bibinfo{year}{2006}.
\newblock \bibinfo{title}{{X-Ray Properties of Black-Hole Binaries}}.
\newblock \bibinfo{journal}{Annu. Rev. Astron. Astrophys.}
  \bibinfo{volume}{44}, \bibinfo{pages}{49--92}.
\newblock \DOIprefix\doi{10.1146/ annurev.astro.44.051905.092532},
  \href{http://arxiv.org/abs/astro-ph/0606352}{\tt arXiv:astro-ph/0606352}.
%Type = Article
\bibitem[{{Samimi} et~al.(1979){Samimi}, {Share}, {Wood}, {Yentis}, {Meekins},
  {Evans}, {Shulman}, {Byram}, {Chubb} and {Friedman}}]{samimi1979gx339}
\bibinfo{author}{{Samimi}, J.}, \bibinfo{author}{{Share}, G.H.},
  \bibinfo{author}{{Wood}, K.}, \bibinfo{author}{{Yentis}, D.},
  \bibinfo{author}{{Meekins}, J.}, \bibinfo{author}{{Evans}, W.D.},
  \bibinfo{author}{{Shulman}, S.}, \bibinfo{author}{{Byram}, E.T.},
  \bibinfo{author}{{Chubb}, T.A.}, \bibinfo{author}{{Friedman}, H.},
  \bibinfo{year}{1979}.
\newblock \bibinfo{title}{{GX339-4 - A new black hole candidate}}.
\newblock \bibinfo{journal}{Nature} \bibinfo{volume}{278},
  \bibinfo{pages}{434--436}.
\newblock \DOIprefix\doi{10.1038/ 278434a0}.
%Type = Article
\bibitem[{{Shakura} and {Sunyaev}(1973)}]{ShaSu1973}
\bibinfo{author}{{Shakura}, N.I.}, \bibinfo{author}{{Sunyaev}, R.A.},
  \bibinfo{year}{1973}.
\newblock \bibinfo{title}{{Black holes in binary systems. Observational
  appearance.}}
\newblock \bibinfo{journal}{Astronomy and Astrophysics} \bibinfo{volume}{24},
  \bibinfo{pages}{337--355}.
%Type = Article
\bibitem[{Shaposhnikov and Titarchuk(2007)}]{shaposhnikov2007determination}
\bibinfo{author}{Shaposhnikov, N.}, \bibinfo{author}{Titarchuk, L.},
  \bibinfo{year}{2007}.
\newblock \bibinfo{title}{Determination of black hole mass in cygnus x-1 by
  scaling of spectral index-qpo frequency correlation}.
\newblock \bibinfo{journal}{The Astrophysical Journal} \bibinfo{volume}{663},
  \bibinfo{pages}{445, ST07}.
%Type = Article
\bibitem[{Shaposhnikov and Titarchuk(2009)}]{shaposhnikov2009determination}
\bibinfo{author}{Shaposhnikov, N.}, \bibinfo{author}{Titarchuk, L.},
  \bibinfo{year}{2009}.
\newblock \bibinfo{title}{Determination of black hole masses in galactic black
  hole binaries using scaling of spectral and variability characteristics}.
\newblock \bibinfo{journal}{The Astrophysical Journal} \bibinfo{volume}{699},
  \bibinfo{pages}{453, ST09}.
%Type = Article
\bibitem[{{Smith} et~al.(2002){Smith}, {Heindl} and {Swank}}]{Smith2002}
\bibinfo{author}{{Smith}, D.M.}, \bibinfo{author}{{Heindl}, W.A.},
  \bibinfo{author}{{Swank}, J.H.}, \bibinfo{year}{2002}.
\newblock \bibinfo{title}{{Two Different Long-Term Behaviors in Black Hole
  Candidates: Evidence for Two Accretion Flows?}}
\newblock \bibinfo{journal}{The Astrophysical Journal} \bibinfo{volume}{569},
  \bibinfo{pages}{362--380}.
\newblock \DOIprefix\doi{10.1086/339167},
  \href{http://arxiv.org/abs/astro-ph/0103304}{\tt arXiv:astro-ph/0103304}.
%Type = Article
\bibitem[{{Sreehari} et~al.(2018){Sreehari}, {Nandi}, {Radhika}, {Iyer} and
  {Mandal}}]{SreeJAA2018}
\bibinfo{author}{{Sreehari}, H.}, \bibinfo{author}{{Nandi}, A.},
  \bibinfo{author}{{Radhika}, D.}, \bibinfo{author}{{Iyer}, N.},
  \bibinfo{author}{{Mandal}, S.}, \bibinfo{year}{2018}.
\newblock \bibinfo{title}{{Observational aspects of outbursting black hole
  sources: Evolution of spectro-temporal features and X-ray variability}}.
\newblock \bibinfo{journal}{Journal of Astrophysics and Astronomy}
  \bibinfo{volume}{39}, \bibinfo{pages}{5}.
\newblock \DOIprefix\doi{10.1007/s12036-018-9510-0},
  \href{http://arxiv.org/abs/1802.05163}{\tt arXiv:1802.05163}.
%Type = Article
\bibitem[{{Sukov{\'a}} and {Janiuk}(2015)}]{Sukova2015}
\bibinfo{author}{{Sukov{\'a}}, P.}, \bibinfo{author}{{Janiuk}, A.},
  \bibinfo{year}{2015}.
\newblock \bibinfo{title}{{Oscillating shocks in the low angular momentum flows
  as a source of variability of accreting black holes}}.
\newblock \bibinfo{journal}{Mon. Not. R. Astron. Soc.} \bibinfo{volume}{447},
  \bibinfo{pages}{1565--1579}.
\newblock \DOIprefix\doi{10.1093/mnras/stu2544},
  \href{http://arxiv.org/abs/1411.7836}{\tt arXiv:1411.7836}.
%Type = Article
\bibitem[{{Sunyaev} and {Titarchuk}(1980)}]{SuTi1980}
\bibinfo{author}{{Sunyaev}, R.A.}, \bibinfo{author}{{Titarchuk}, L.G.},
  \bibinfo{year}{1980}.
\newblock \bibinfo{title}{{Comptonization of X-rays in plasma clouds - Typical
  radiation spectra}}.
\newblock \bibinfo{journal}{Astronomy and Astrophysics} \bibinfo{volume}{86},
  \bibinfo{pages}{121--138}.
%Type = Article
\bibitem[{{Svensson} and {Zdziarski}(1994)}]{Svensson1994}
\bibinfo{author}{{Svensson}, R.}, \bibinfo{author}{{Zdziarski}, A.A.},
  \bibinfo{year}{1994}.
\newblock \bibinfo{title}{{Black hole accretion disks with coronae}}.
\newblock \bibinfo{journal}{The Astrophysical Journal} \bibinfo{volume}{436},
  \bibinfo{pages}{599--606}.
\newblock \DOIprefix\doi{10.1086/174934}.
%Type = Article
\bibitem[{{Tetarenko} et~al.(2016){Tetarenko}, {Sivakoff}, {Heinke} and
  {Gladstone}}]{Tetarenko2016}
\bibinfo{author}{{Tetarenko}, B.E.}, \bibinfo{author}{{Sivakoff}, G.R.},
  \bibinfo{author}{{Heinke}, C.O.}, \bibinfo{author}{{Gladstone}, J.C.},
  \bibinfo{year}{2016}.
\newblock \bibinfo{title}{{WATCHDOG: A Comprehensive All-sky Database of
  Galactic Black Hole X-ray Binaries}}.
\newblock \bibinfo{journal}{The Astrophysical Journal Supplement}
  \bibinfo{volume}{222}, \bibinfo{pages}{15}.
\newblock \DOIprefix\doi{10.3847/0067-0049/222/2/15},
  \href{http://arxiv.org/abs/1512.00778}{\tt arXiv:1512.00778}.
%Type = Article
\bibitem[{{Titarchuk} and {Fiorito}(2004)}]{TF04}
\bibinfo{author}{{Titarchuk}, L.}, \bibinfo{author}{{Fiorito}, R.},
  \bibinfo{year}{2004}.
\newblock \bibinfo{title}{{Spectral Index and Quasi-Periodic Oscillation
  Frequency Correlation in Black Hole Sources: Observational Evidence of Two
  Phases and Phase Transition in Black Holes}}.
\newblock \bibinfo{journal}{The Astrophysical Journal} \bibinfo{volume}{612},
  \bibinfo{pages}{988--999}.
\newblock \DOIprefix\doi{10.1086/422573},
  \href{http://arxiv.org/abs/astro-ph/0405360}{\tt arXiv:astro-ph/0405360}.
%Type = Article
\bibitem[{{Titarchuk} and {Lyubarskij}(1995)}]{TitLyu1995}
\bibinfo{author}{{Titarchuk}, L.}, \bibinfo{author}{{Lyubarskij}, Y.},
  \bibinfo{year}{1995}.
\newblock \bibinfo{title}{{Power-Law Spectra as a Result of Comptonization of
  the Soft Radiation in a Plasma Cloud}}.
\newblock \bibinfo{journal}{The Astrophysical Journal} \bibinfo{volume}{450},
  \bibinfo{pages}{876}.
\newblock \DOIprefix\doi{10.1086/176191}.
%Type = Article
\bibitem[{{Wu} et~al.(2002){Wu}, {Soria}, {Campbell-Wilson}, {Hannikainen},
  {Harmon}, {Hunstead}, {Johnston}, {McCollough} and {McIntyre}}]{Wu2002}
\bibinfo{author}{{Wu}, K.}, \bibinfo{author}{{Soria}, R.},
  \bibinfo{author}{{Campbell-Wilson}, D.}, \bibinfo{author}{{Hannikainen}, D.},
  \bibinfo{author}{{Harmon}, B.A.}, \bibinfo{author}{{Hunstead}, R.},
  \bibinfo{author}{{Johnston}, H.}, \bibinfo{author}{{McCollough}, M.},
  \bibinfo{author}{{McIntyre}, V.}, \bibinfo{year}{2002}.
\newblock \bibinfo{title}{{The 1998 Outburst of XTE J1550-564: A Model Based on
  Multiwavelength Observations}}.
\newblock \bibinfo{journal}{The Astrophysical Journal} \bibinfo{volume}{565},
  \bibinfo{pages}{1161--1168}.
\newblock \DOIprefix\doi{10.1086/324328},
  \href{http://arxiv.org/abs/astro-ph/0109222}{\tt arXiv:astro-ph/0109222}.
%Type = Article
\bibitem[{{Zycki} et~al.(1995){Zycki}, {Collin-Souffrin} and
  {Czerny}}]{Zycki1995}
\bibinfo{author}{{Zycki}, P.T.}, \bibinfo{author}{{Collin-Souffrin}, S.},
  \bibinfo{author}{{Czerny}, B.}, \bibinfo{year}{1995}.
\newblock \bibinfo{title}{{Accretion Discs with Accreting Coronae in Active
  Galactic Nuclei - Part One - Solutions in Hydrostatic Equilibrium}}.
\newblock \bibinfo{journal}{Mon. Not. R. Astron. Soc.} \bibinfo{volume}{277},
  \bibinfo{pages}{70}.
\newblock \DOIprefix\doi{10.1093/mnras/277.1.70},
  \href{http://arxiv.org/abs/astro-ph/9505045}{\tt arXiv:astro-ph/9505045}.

\end{thebibliography}
%\clearpage

\end{document}